\documentclass[a4paper,aip,jcp,floatfix,unsortedaddress,superscriptaddress,reprint]{revtex4-1}
\usepackage[utf8]{inputenc}
\usepackage{amsmath,bm}
\usepackage{graphicx}
\usepackage{color}
\usepackage{verbatim}
\usepackage[position=t,singlelinecheck=false,caption=false]{subfig}

\newcommand{\celsius}{\,^{\circ}{\rm C}}
\newcommand{\angstrom}{\textup{\AA}}
\newcommand{\sodium}{[Na$^+$]}

\newcommand*{\plimsoll}{{\ensuremath{-\kern-4pt{\ominus}\kern-4pt-}}}

\begin{document}

\title{A nucleotide-level coarse-grained model of RNA}

\author{Petr \v{S}ulc}
\affiliation{Rudolf Peierls Centre for Theoretical Physics, 
 University of Oxford, 1 Keble Road, Oxford, OX1 3NP, United Kingdom}

\author{Flavio Romano}
\affiliation{Physical and Theoretical Chemistry Laboratory, 
 Department of Chemistry, University of Oxford, South Parks Road, 
 Oxford, OX1 3QZ, United Kingdom}

\author{Thomas E.~Ouldridge}
\affiliation{Rudolf Peierls Centre for Theoretical Physics, 
 University of Oxford, 1 Keble Road, Oxford, OX1 3NP, United Kingdom}

\author{Jonathan~P.~K. Doye}
\affiliation{Physical and Theoretical Chemistry Laboratory, 
 Department of Chemistry, University of Oxford, South Parks Road, 
 Oxford, OX1 3QZ, United Kingdom}

\author{Ard~A. Louis}
\affiliation{Rudolf Peierls Centre for Theoretical Physics, 
 University of Oxford, 1 Keble Road, Oxford, OX1 3NP, United Kingdom}

\date{\today}

\begin{abstract}

We present a new, nucleotide-level model for RNA, oxRNA, based on the coarse-graining methodology recently developed for the oxDNA model of DNA.
The model is designed to reproduce structural, mechanical and thermodynamic properties of RNA, and the coarse-graining level aims to retain the
relevant physics for RNA hybridization and the structure of single- and
double-stranded RNA.  In order to explore its strengths and weaknesses,  we test the model in a range of nanotechnological and biological settings.  Applications explored include the folding thermodynamics of a pseudoknot, the formation of a kissing loop complex, the structure of a hexagonal RNA nanoring, and the unzipping of a hairpin motif.  We argue that the model can be used for efficient simulations of the structure of systems with thousands of base pairs, and for the assembly of systems of up to
hundreds of base pairs. The source code implementing the model is released for
public use.

\end{abstract}

\maketitle
\section{Introduction}
RNA (ribonucleic acid) strands are polymers that play crucial cellular roles in gene
expression, translation and regulation.\cite{Elliott11} RNA is
composed of units called nucleotides, which consist of a phosphate and ribose
sugar backbone to which one of four different bases is attached: adenine (A),
guanine (G), cytosine (C) or uracil (U).  DNA has a similar chemical structure,
but thymine (T) is present in place of uracil and the backbone includes deoxyribose sugars instead of ribose sugars. 
 Both RNA and DNA can form double-helical
molecules, stabilized by hydrogen bonds between complementary
Watson-Crick base pairs: AU (AT in case of DNA) and GC. For RNA, wobble base
pairs (GU) can  also stabilize the duplex form and are commonly found.
While DNA typically forms a B-helix, double-stranded RNA is typically found in a related, but different A-helix geometry.\cite{Saenger1984}  

DNA's role in biological systems is to store genetic information and it is most often found in its double-helical
form. By contrast, most naturally occurring RNA molecules are single-stranded and fold into
complex structures that contain double-helical segments as well as loops,
junctions and numerous tertiary structure interactions that further stabilize
the molecule. The lengths of RNA strands found in living cells can range from a few
nucleotides to several thousands or even tens of thousands.\cite{rnadb}

DNA nanotechnology\cite{seeman2010nanomaterials} uses
DNA molecules to create nanoscale structures and active devices. Examples of experimentally realized systems include
DNA motors,\cite{Bath2005,Bath2009} self-assembled nanostructures such as DNA origamis\cite{Rothemund06} and the 
use of DNA strands for computation.\cite{Adleman9} Since RNA molecules are more
difficult to handle and preserve than DNA in experimental settings, developing
RNA-based nanodevices has been a more challenging task. However, the emerging
field of RNA nanotechnology\cite{guo2010} offers promising applications of
RNA-based nanodevices both {\em in vitro} and {\em in vivo}. The general design
principles from DNA nanotechnology can be applied to RNA, but because
specific RNA sequences form functional structures that are known to interact with proteins
and other RNA molecules in the cell, RNA molecules would be the natural choice for
nanotechnology devices operating inside cells.\cite{benenson2009}
 Successful examples of
experimentally realized RNA nanotechnology include self-assembling RNA nanocubes\cite{afonin2010} and RNA
tiles.\cite{chworos2004} Recently, an RNA strand displacement reaction cascade was proposed that could be used for conditional gene silencing inside the cell.\cite{lisapeng} 

Because of its importance for biological systems, RNA has been 
the subject of intensive study. In addition to numerous experimental studies, a range of theoretical and computational methods have been applied
to the study of its properties. Currently, there are multiple tools and computational approaches that can
describe RNA structure at various levels of detail.\cite{laing2010computational,laing2011computational}

In an important series of
works\cite{serra1995, xia1998thermodynamic, mathews2004incorporating,
mathews1999expanded, walter1994sequence, walter1994coaxial, lu2006set} the
thermodynamics of RNA secondary structure (i.e., the list of
base pairs in the folded state of RNA) was characterized in terms of a nearest-neighbor model for calculating the free energies of various secondary-structure motifs. The nearest-neighbor model is the basis of various tools for the prediction of the secondary structure. Such tools typically use dynamic programming
approaches to find the secondary structure with minimal free
energy.\cite{nupack,hofacker1994fast,reuter2010rnastructure}   Furthermore, some tools have been extended by adding simple kinetic descriptions to the nearest-neighbor thermodynamics,  allowing folding transitions to be modeled.\cite{flamm2000rna,kinefold}  Although these methods are typically very fast,  the fundamentally discrete nature of the description and the lack of structural and mechanical detail places a limit on what they can treat. 

At the highest level of detail, quantum chemistry methods have been employed
to study the interactions between nucleotides.\cite{Sponer04,Svozil10} Due to
the complexity of the calculations, such methods remain limited to interactions
between nearest-neighbor base pairs in vacuum. Fully atomistic
molecular dynamics simulation packages such as Amber\cite{cornell95} or
CHARMM\cite{CHARMM} include the solvent, and use classical effective interaction potentials between atoms to represent systems 
with high resolution. However, the study of rare
events such as the formation or breaking of individual base pairs remains a very challenging
task. Simulations of
several folding pathways of a short RNA
hairpin\cite{bowman2008structural} and tetraloops\cite{kuhrova2013computer} provide examples of the limit of what is currently possible.  At present, atomistic molecular dynamics simulations can attain timescales of the
order of $\mu$s.   Moreover, while the forcefields are improving, they are still under development so that different versions can generate different behavior.\cite{Krepl2012,Yoo2013,guy2012single}
A recently developed approach\cite{minary2008dynamical} combines fully atomistic representation with hierarchical Monte Carlo sampling, where different series of moves are used to move whole sections of a molecule (such as all atoms contained in one stem) at once. Such methods have for instance been used to study the effects of mutations in a sequence on the conformational freedom of a tRNA molecule and of a nanosquare composed of four tRNAs.\cite{minary12}

In order to access longer timescales relevant to rare events, such as the
breaking of base pairs or the formation of large structures, one needs to use a
more coarse-grained description. In this approach, atoms are incorporated into a reduced set of degrees of freedom that 
experience effective interactions.  Solvent molecules are often integrated out.
Such models always present a compromise between accuracy, efficiency and the level of detail, which determines their scope. Several
coarse-grained RNA models have been developed in recent years.\cite{jonikas2009coarse,xiamodel,parisien2008mc,das2010atomic,paliy2010coarse,Pasquali2010,Cragnolini13,ding2008ab,Hyeon2005,denesyuk2013coarse,hyeon2006pathways,cao2005predicting,jost2010prediction} 

Knowledge-based coarse-graining uses the information extracted from experimentally determined crystal structures 
to develop potentials, usually with the goal to predict the folded structure for an RNA sequence, either {\it de novo} or with some additional input of data from the user.
An example of such an approach is the NAST model \cite{jonikas2009coarse} which represents each nucleotide 
as a single pseudoatom in the simulation and uses a statistical potential, inferred from known structures of RNA molecules, that depends on the distances and angles between the nucleotides.
This model requires the secondary structure and tertiary contacts of the final folded RNA structure as an input for the folding simulation. It has been used to study RNA structures of up to 158 nucleotides. While it was able to reproduce the structures of folded RNA, it was not parametrized to reproduce their thermodynamic properties.

Similarly, the model of Xia {\it et al.},\cite{xiamodel} which uses 6 beads to represent a nucleotide, has interactions parametrized to reproduce known RNA structures. Xia {\it et al.}~were able to predict the tertiary structure of several RNA molecules of lengths up to 122 nucleotides by using simulated annealing to attempt to find the
global potential energy minimum of the structures in their coarse-grained model. The resulting structures were then refined by a simulation with a fully atomic representation. The thermodynamic properties of the coarse-grained model were not reported.

Finally, some knowledge-based methods combine together various structural motifs from database of experimentally determined RNA structures to predict folded RNA structure for a given sequence. The algorithms match fragments with a particular section in the RNA strand. For example, Parisien and Major\cite{parisien2008mc} used such an approach to predict the secondary and tertiary structure of RNA strands with up to 50 nucleotides. The FARFAR method\cite{das2010atomic} further uses sampling with a fully atomistic representation of the respective RNA residues in order to obtain the final structure. It successfully predicted {\it de novo} folded structures for RNA sequences of size up to 20 nucleotides.

An alternative approach for model development is to use fully atomistic simulations to parametrize the effective interactions between 
coarse-grained representations of groups of atoms. Such an approach was adopted, for example, by Paliy {\it et al.},\cite{paliy2010coarse} who presented different levels of coarse-graining,
using either one or three beads per nucleotide. The interactions between beads were fitted to reproduce the probability distribution of their mutual orientations and distances, calculated from from a simulation with a fully atomistic representation. The authors were then able to simulate the conformations of an RNA nanoring structure which consisted of 330 nucleotides.  

The HiRe-RNA model\cite{Pasquali2010,Cragnolini13} represents each 
nucleotide as 6 or 7 beads with empirically chosen interactions based on a combination of atomistic simulations and known structures. It reproduces the structure of RNA duplexes and was used to simulate the
association and dissociation of small oligonucleotides (16 base pairs). The model further allows the reconstruction of a fully atomistic representation of an RNA molecule from its coarse-grained representation. The model was also used to study some transitions in RNA, although a direct link between the parameters and experimental melting temperatures has not yet been made.  

The above mentioned models were parametrized to structure, either through comparison to experiment, or to atomistic simulations from which thermodynamic quantities are hard to extract.    While that is useful for the structure of folded RNA complexes, it makes it hard to compare with available experimental data on RNA thermodynamics, or to simulate reactions involving multiple RNA strands.  The next set of models do include explicit thermodynamic information in their parametrization.

The coarse-grained model of Ding {\it et al}.\cite{ding2008ab}~uses three beads
(sugar, phosphate and base) to represent each nucleotide and has been used to
study the folding of various RNA structures, including tRNA and pseudoknots, of sizes up to 100 nucleotides. The parametrization of the interactions combines a knowledge-based approach with a parametrization of the interaction strengths to the free energies of base pairs taken from the nearest-neighbor model. Their simulation algorithm furthermore takes into account explicitly the free-energy cost for closing a loop as predicted by the nearest-neighbor model. This added free-energy contribution does not come from the model's interactions and hence ties the use of the model to this particular simulation algorithm. 

The model of Hyeon and Thirumalai\cite{Hyeon2005} also uses three beads per nucleotide. Its interaction strengths are based on nearest-neighbor model parameters. The model was used to study mechanical unfolding of hairpins. Recently, the model was extended by Denesyuk and Thirumalai\cite{denesyuk2013coarse} with interactions
 parametrized using thermodynamic data from pseudoknot and hairpin melting experiments combined with the free energies in the nearest-neighbor model for RNA thermodynamics.\cite{mathews1999expanded} The new model has been used to study the thermodynamics of folding of a 34-nucleotide pseudoknot. The model also includes explicit electrostatic interactions and can also represent tertiary structure contacts such as hydrogen bonds in non-canonical base pairs. We note that Hyeon {\em et al}.~also developed the SOP model for RNA,\cite{hyeon2006pathways} which only uses one site per nucleotide, to study larger systems. The interactions in the model were set to a given energy scale and were not compared with RNA thermodynamics. The SOP model was used to study the mechanical unfolding of a 421-base ribozyme.\cite{hyeon2006pathways} 

The nearest-neighbor model was also used to parametrize the lattice-based model of Cao and Chen\cite{cao2005predicting} which represents the conformations of RNA as a self-avoiding walk on a 3D-lattice. This model was used to compute the heat capacities from  partition functions for different mutants of the so-called 72 RNA structure, which were found to be in good agreement with experimental measurements. It was further used to study the free-energy landscape at different temperatures for a 76-nucleotide P5abc RNA structure. 

Finally, the lattice model of Jost and Everaers\cite{jost2010prediction} is parametrized to reproduce the nearest-neighbor model thermodynamics. The parametrization was verified by studying thermodynamics of ten RNA hairpins and an ensemble of structures with varying internal loop sizes. The model was then used to study folding pathways of a 76 nucleotide long tRNA and a pseudoknot. While lattice based models allow for an efficient sampling of the possible conformations, the structural description of the RNA is necessarily limited by the requirement that it is placed on a lattice. 

Most of the existing coarse-grained RNA models are aimed at the correct prediction of the most probable folded structure for a given RNA sequence. In these cases, the thermodynamics of RNA was either not considered, or was used to guide parameter choice which was then tested on a few selected systems. Of the described models, the most detailed verification of the  thermodynamics was for the model of Jost and Everaers. 
We further note that mechanical properties have not been reported for any of these models.  

Here we propose a new off-lattice coarse-grained RNA model, oxRNA, that follows
the coarse-graining approach developed for the DNA model
oxDNA.\cite{oxDNA,Ouldridge2011,Ouldridge_thesis} Given that oxDNA has been successfully
used to model DNA nanotechnological systems, such as 
motors,\cite{ouldridge2013optimizing,sulc2012simulating} 
tweezers,\cite{Ouldridge_tweezers_2010} kissing
hairpins,\cite{Romano12a,oxDNA} strand
displacement,\cite{srinivas2013biophysics} as well as for biophysical applications including cruciforms,\cite{Matek} the
pulling of single\cite{oxDNA} and double-stranded DNA\cite{romano2013coarse} and the
hybridization of short oligonucleotides,\cite{ouldridge2013dna} our goal is to derive a model of similar applicability for RNA.
 We aim to capture basic RNA structure, mechanics and thermodynamics in as minimalistic a description as possible. We replace each RNA nucleotide by a single
rigid body with multiple interaction sites. 
The interactions between rigid
bodies are parametrized to allow an A-helix to form from two single strands and to
reproduce RNA thermodynamics.
The resultant model goes beyond nearest-neighbor thermodynamics because it has the ability to capture topological, mechanical and spatial effects and allows for the study of kinetic properties of various processes within a molecular dynamics simulation framework.  
The model uses only pairwise interactions to facilitate the use of cluster Monte Carlo algorithms for simulations. The simple representation, one rigid body per nucleotide, allows for efficient simulation of structures of sizes up to several hundred nucleotides on a single CPU as well as of rare events such as the dissociation or the formation of a double helix.

This paper is organized as follows.
In Section \ref{sec_model_and_par} we present our coarse-grained model and
its parametrization. In Section \ref{sec_model_proper} we study the thermodynamic, structural and mechanical properties of the model.  We then illustrate the utility of the model through applications to pseudoknot thermodynamics, hairpin unzipping and kissing
hairpins in Section \ref{sec_examples}. The detailed description of the interaction potentials in our model is provided in the Supplementary Material.

\section{The RNA model and its parametrization}
\label{sec_model_and_par}
In this section we describe the parametrization of the oxRNA model to reproduce the RNA thermodynamics of the nearest-neighbor model, which we briefly outline in Section \ref{sec_nnmodel}.

\subsection{RNA thermodynamics and the nearest-neighbor model}
\label{sec_nnmodel}
In an extensive series of investigations,\cite{mathews1999expanded,xia1998thermodynamic,lu2006set,mathews2004incorporating} Turner {\it et al.}~parametrized a nearest-neighbor model (hereafter referred to as the NN-model) to describe the thermodynamics of RNA duplex and hairpin formation, which is widely used in RNA secondary structure 
prediction.\cite{nupack,unafold,kinefold,ViennaRNA,bellaousov2013rnastructure} The model treats RNA at the level of secondary structure, estimating enthalpic and entropic contributions to the stability from each pair of consecutive base pairs (bp) in a structure 
and including corrections for end effects and enclosed loops of unpaired bases. The parametrization used melting experiments of short duplexes and hairpins at $1\,{\rm M}$ \sodium. The results were fitted using a two-state assumption in which RNA either adopts the fully-formed structure or is completely disordered. 
The yield of the duplexes is then given by 
\begin{equation}
\label{eq_duplex_yield}
\frac{[AB]}{[A][B]} = \exp(-\Delta G^\plimsoll (T) / RT),
\end{equation}
where  $[AB]$ is the molar concentration of the duplexes, and $[A]$ and $[B]$
the concentrations of the single strands. $\Delta G^\plimsoll (T) = \Delta H^\plimsoll - T \Delta S^\plimsoll$ is
the standard Gibbs free-energy change that is given by the NN-model, where $\Delta H^\plimsoll$ and $\Delta S^\plimsoll$
are calculated from the base-pair step contributions. 
The concentrations of reactants are specified relative to 1M and $\Delta G^\plimsoll$ is calculated for a system where all reactants have a concentration of 1M.

Similarly, the yields of hairpins are
\begin{equation}
\frac{[C]}{[O]} = \exp(-\Delta G^\plimsoll(T) / RT),
\end{equation}
where $[C]$ is the concentration of closed strands, and $[O]$ is the
concentration of open strands. 

The melting temperature of a duplex, at a given strand concentration, is
defined as the temperature at which half of the duplexes present in the solution
are dissociated into single strands. Similarly, the melting temperature of a
hairpin is defined as the temperature at which the strand has a 50\% probability of
being open.

The NN-model has been shown to reproduce the melting temperatures of RNA
oligonucleotides with Watson-Crick base-pairing with $1.3\,^{\circ}{\rm C}$
accuracy.\cite{xia1998thermodynamic} In our work, we will treat the NN-model as
an accurate fit to the melting data and use its melting temperatures
predictions for fitting the oxRNA model. 
The average accuracy (the fraction of correctly predicted base pairs in known secondary structures) of the NN-model is reported to be 73\% for a database of strands that has in total 150\,000 nucleotides, containing domains of up to 700 nucleotides.\cite{mathews2004incorporating}

\subsection{The Representation}

\begin{figure}[tb]
\includegraphics[width=0.5\textwidth]{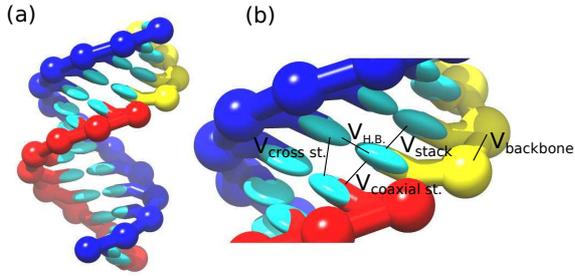}
\centering
\caption{A schematic representation of (a) an A-RNA helix as represented by the model and of (b) the attractive interaction in oxRNA. The nucleotides can also interact with
excluded-volume interactions.}
\label{fig_potentials}
\end{figure}

OxRNA uses a single rigid body with multiple
interaction sites to represent a nucleotide. Each rigid body has a backbone, 3$^\prime$-stacking, 5$^\prime$-stacking, cross-stacking, and hydrogen-bonding interaction sites. The detailed description of the representation of a nucleotide is given in the Supplementary Material (Fig.~\ref{fig_nuc}). In the figures we use a schematic ellipsoid to represent the stacking and hydrogen-bonding sites as this allows the orientation of the base to be clearly seen.
The potentials between the nucleotides are effective interactions that are designed to capture the overall thermodynamic and structural consequence of the base-pairing and stacking interactions, rather
than directly representing the microscopic contributions such as electrostatics, dispersion, exchange repulsion and hydrophobicity.

We choose the functional forms of our coarse-grained interactions to reproduce
directly experimentally measured properties of RNA. For these reasons, we label
our coarse-graining approach ``top-down'', as opposed to a ``bottom-up'' approach
which starts from a more detailed description of the system. We point out that
any coarse-grained interaction is actually a free-energy for the real
system, rather than a potential energy, and therefore it is in principle
state-point dependent. So it should come to no surprise that our potential
contains an explicit dependence on the temperature, although for simplicity we
try to limit this as much as possible and only introduce it in one of the
interaction terms ($V_{\rm{stack}}$, as we will discuss later).
Our coarse-graining aims to retain the relevant geometric degrees of freedom in order to still correctly capture the relative entropies of different states, despite not having temperature dependence in most of the interaction potentials.\cite{Doye13}

The potential energy of the model is 
\begin{eqnarray}
\label{eq_potential}
 V_{\rm oxRNA} &=& \sum_{\left\langle ij \right\rangle} \left( V_{\rm{backbone}} + V_{\rm{stack}} +
V^{'}_{\rm{exc}} \right) +  \\
   &+& \sum_{i,j \notin {\left\langle ij \right\rangle}} \left( V_{\rm{H.B.}} +  V_{\rm{cross~st.}}  + V_{\rm{exc}}  + V_{\rm{coaxial~st.}} \right) , \nonumber 
\end{eqnarray}
where the first sum is taken over all the nucleotides that are neighbors along an
RNA strand and the second sum is taken over all the non-nearest-neighbor pairs
of nucleotides. All potentials are two-body potentials.  There is a maximum distance beyond which all potentials are zero (with the exception of $V_{\rm{backbone}}$ which diverges to infinity as the distance between adjacent backbone sites approaches its maximum value). The interactions are schematically shown in Fig.~\ref{fig_potentials}. We discuss briefly the potentials here
while the detailed description is given in the Supplementary Material \ref{Functional_forms}.

The backbone interaction, $V_{\rm{backbone}}$, is an isotropic FENE (finitely-extensible nonlinear elastic) potential
and depends only on the distance
between the backbone sites of the two adjacent nucleotides. This potential is
used to mimic the covalent bonds in the RNA backbone that constrain this intramolecular distance. The nucleotides also
have repulsive excluded-volume interactions $V_{\rm{exc}}$ and $V^{'}_{\rm{exc}}$ that depend on the distance between the interaction
sites, namely the backbone-backbone, stacking-stacking and stacking-backbone distances.
The excluded-volume interactions ensure that strands cannot overlap, or pass through each
other in a dynamical simulation.

The duplex is stabilized by hydrogen bonding ($V_{\rm{H.B.}}$), stacking
($V_{\rm{stack}}$) and  cross-stacking ($V_{\rm{cross~st.}}$) interactions.
These potentials are highly anisotropic and depend on the distance between the relevant
interaction sites as well as the mutual orientations of the nucleotides.  The anisotropic potentials are
of the form 
\begin{eqnarray}
 \label{eq_potential_type}
 V_{\rm{H.B.}} &=& \alpha_{ij} f_{\rm{H.B.}} \left(\mathbf{r}_{ij}, \mathbf{\Omega}_i, \mathbf{\Omega}_j \right) \\
 V_{\rm{stack}} &=& \eta_{ij} (1 + \kappa \, k_{\rm B} T) f_{\rm{stack}} \left(\mathbf{r}_{ij}, \mathbf{\Omega}_i, \mathbf{\Omega}_j \right) \label{eq_bt}  \\
 V_{\rm{cross~st.}} &=& \gamma f_{\rm{cross~st.}} \left(\mathbf{r}_{ij}, \mathbf{\Omega}_i, \mathbf{\Omega}_j \right) \\
 V_{\rm{coaxial~st.}} &=& \mu f_{\rm{coaxial~st.}} \left(\mathbf{r}_{ij}, \mathbf{\Omega}_i, \mathbf{\Omega}_j \right)
\end{eqnarray}
 where the functions $f$ are a product of multiple terms, one of which depends on the
distance between the relevant interaction sites and the remaining
are angular modulation functions that
are equal to one if the relevant
angles between the nucleotides correspond to the minimum potential energy configuration, and
smoothly go to zero as they depart from these values. The set of angles is different
for each potential and includes angles between intersite vectors and orientations $\mathbf{\Omega}_i$ and $\mathbf{\Omega}_j$ of the nucleotides. The constant prefactors $\alpha_{ij},\eta_{ij}, \gamma$, and $\mu$
set the strength of the interactions, with $\alpha_{ij},\eta_{ij}$ being dependent on the nucleotides involved. 

The hydrogen-bonding term $V_{\rm{H.B.}}$ is designed to capture the
duplex stabilizing interactions between Watson-Crick and wobble base pairs. The
potential reaches its minimum when two complementary nucleotides (AU, GC or
GU) are coplanar, directly opposite and antiparallel with respect to each
other and at the right distance. 

The stacking interaction $V_{\rm{stack}}$ mimics the favorable interaction between adjacent bases, which results from a combination of hydrophobic, electrostatic and dispersion effects.
It acts only between nearest-neighbor nucleotides and its
strength depends on both the distance between the respective 3$^{\prime}$ and 5$^{\prime}$ stacking
sites of the nucleotides as well as their mutual orientations.
It also depends on the vector between the backbone interaction sites in a way that
ensures the inclination of the bases in the duplex structure matches that for A-RNA. We note that the nucleotides can also
interact via the stacking interaction when they are in the single-stranded state. To
ensure the right-handedness of the RNA helix in the duplex state, the stacking
interaction has an additional modulation term that is equal to one if the
nucleotides adopt a right-handed conformation and goes smoothly to zero in the left-handed
conformation.

Similarly to oxDNA, the interaction strength of the stacking potential has a
temperature-dependent contribution (the term $\kappa\, k_{\rm B} T$ in Eq.~\ref{eq_bt}). This term was introduced in oxDNA in order to correctly reproduce the thermodynamics
of the stacking transition. We also found that retaining this temperature dependence enables oxRNA to reproduce more accurately the widths of the melting transitions,
which are discussed in more detail in Section \ref{sec_parametrization}.

The cross-stacking potential, $V_{\rm{cross~st.}}$, is designed to capture the
interactions between diagonally opposite bases in a duplex and has its minimum when the distance and mutual orientation
between nucleotides correspond to the arrangement of a nucleotide and a
3$^\prime$ neighbor of the directly opposite nucleotide in a A-helix. This interaction has been parametrized to capture the stabilization of an RNA duplex by a 3$^{\prime}$ overhang.\cite{mathews1999expanded} OxRNA does not include any
interaction with the
5$^\prime$ neighbor of the directly opposite nucleotide, as 5$^\prime$ overhangs are significantly less
stabilizing than 3$^\prime$
overhangs.\cite{mathews1999expanded}  

Finally, the coaxial stacking potential $V_{\rm{coaxial~st.}}$  represents the
stacking interaction between nucleotides that are not nearest-neighbors on the
same strand.

We note that, although our model does not include an explicit term for electrostatic
interactions between phosphates, these interactions are effectively incorporated into the backbone repulsion. We chose to parametrize our model to the experimental data at $1\,\rm{M}$
\sodium, where the electrostatic interaction is highly screened, making our approach reasonable. Furthermore, we are only able to capture those tertiary structure motifs that involve Watson-Crick and wobble base pairing or stacking, such as kissing hairpins or coaxial stacking of helices. In particular, oxRNA does not include  
Hoogsteen or sugar-edge hydrogen-bonded base pairs, or ribose zippers (interactions involving the 2$^\prime$-OH
group on the ribose sugar). In principle these interactions could be included, but for this version of the model we chose not to as there are no systematic thermodynamic data to which
we could parametrize the relevant interaction strengths.  

While the strengths of the hydrogen-bonding and stacking interactions depend on the identities of the interacting nucleotides, as in oxDNA, all nucleotides in oxRNA have the same size and shape. Therefore we do not expect oxRNA to capture detailed sequence-dependent structure of the A-helix.

The positions of the minima in the potential functions have been selected so that the model reproduces the structure of the A-RNA double helix, which RNA duplexes have been shown to adopt\cite{Saenger1984,Elliott11} and which we describe in more detail in Section \ref{sec_structure}. The widths of the potential functions and the strengths of the interaction potentials were parametrized to reproduce RNA thermodynamics as described in Section \ref{sec_parametrization}.

\subsection{Simulation methods}
\subsubsection{Algorithms}
For the majority of our simulations, unless noted otherwise, we use the Virtual Move
Monte Carlo algorithm (VMMC) (specifically the variant described in the Appendix of Ref.~\onlinecite{Whitelam2009}) to obtain the thermodynamics of oxRNA. VMMC is
a cluster-move Monte Carlo algorithm that has previously been used in many
applications for the oxDNA model.\cite{Doye13} We also implemented the oxRNA forcefield in a  molecular
dynamics (MD) code with the choice of a Langevin\cite{schlick2010molecular} or an Andersen-like\cite{Russo09} thermostat. For the MD simulations in this work, we used the Andersen-like thermostat.

In our simulations, we often combine the VMMC algorithm with the umbrella
sampling method\cite{Torrie1977} in order to help systems overcome 
free-energy barriers. This technique involves splitting the configuration space
of the system into regions using order parameters, and then
assigning weights to each state so defined. In our case, the order parameter is
typically the number of base pairs in a given system. 
The probability of
accepting a move to a state with a different weight is then scaled by the ratio
of the weights of the two respective states. By assigning larger weights to less
probable states (such as states with only a few base pairs), one can achieve an
efficient sampling of all values of the order parameter. When extracting the
occupancy probabilities for different states from our simulation, we unbias them
by rescaling by the inverse of the weight of each state. The weights are
typically chosen by experience and then adapted iteratively by hand, until
a selection of weights that samples efficiently all relevant states of
the system is obtained.

\subsubsection{Calculation of melting temperatures}
\label{sec_rewei}
When calculating the free energy of a system as a function of an order parameter
(typically the number of base pairs between two strands or, for example, the
number of base pairs in a hairpin stem), we run multiple simulations that sample
all relevant states, aided by using umbrella sampling. We then calculate $p_i$, the
unbiased probabilities of the system having a particular value $i$ of the order parameter,
and take the free energy of such a state to be $F_i = -k_{\rm B} T \ln \left(
p_i\right) + c$, where $k_{\rm B}$ is the Boltzmann constant and $c$ is a constant. We are interested only in differences in free energies between different states and hence $c$ can be set to an arbitrary value.   

When calculating the melting temperature of a duplex, we simulate a system of
two complementary strands and calculate the ratio $\Phi$ of the times that the
system spends in a duplex state and in a dissociated state. At the melting
temperature $T_m$, $\Phi = 2$ for heterodimers, accounting
for finite-size effects that come from simulating only two strands, as explained in Refs.~\onlinecite{Ouldridge_bulk_2010}, \onlinecite{ouldridge2012inferring} and
\onlinecite{de2011determining}. For transitions involving single strands, such as hairpin formation, the ratio $\Phi$ is the time spent in closed states (i.e.\ states with at least one base pair in the stem) divided by the time spent in open states. In this case, $\Phi=1$ at the melting temperature because the finite size effects mentioned above are irrelevant for unimolecular assembly. 
We note that we consider the system to be 
in a duplex (hairpin) form if the complementary strands (stems) have one or more base pairs present.

Using the histogram reweighting method, the states generated during the umbrella sampling simulations can be used to calculate the melting temperatures for a model with the same functional forms of the potential but with different interaction strengths. Previously, we used this approach for the parametrization of the oxDNA model and the method is described in detail in Ref.~\onlinecite{oxDNA}.
The reweighting method recalculates the ratio $\Phi$ for temperature $T$ and a given set of interaction strengths by counting the contribution to the bonded or unbonded state of each sampled state with an additional weight factor  
$$w = \exp \left(\frac{V_0(T_0)}{k_{\rm B} T_0} - \frac{ V(T) }{k_{\rm B} T} \right) $$
where $V_0$ is the potential energy of the state generated in the simulation with the original set of parameters at temperature $T_0$, and $V(T)$ is the potential energy of the same state recalculated for the interaction strengths for which we want to find the ratio $\Phi$ at temperature $T$. The reweighting method allows us to extract $\Phi$ for a given set of interaction strengths at a range of temperatures which then can be interpolated to find the melting temperature. This method assumes that the ensemble of states generated with the original set of parameters is representative of the states
that the system would visit in a simulation with the new parameters.
Using this approach it is possible to calculate the melting temperatures for
thousands of different sets of interaction constants
in an hour of CPU time without the necessity of rerunning the umbrella sampling simulation, which, by contrast, can take several days on a single CPU for each sequence to calculate the melting temperature to within $1\celsius$ accuracy.

\subsection{Parametrization of the model}
\label{sec_parametrization}

The anisotropic potentials $V_{\rm{stack}}$, $V_{\rm{H.B.}}$ and
$V_{\rm{cross~st.}}$ have interaction strengths of the form $\eta_{ij} ( 1 + \kappa \, k_{\rm B} T) $,
$\alpha_{ij}$ and $\gamma$ respectively, where the stacking interaction strength
depends also on the simulation temperature $T$ and $i$ and $j$ correspond to the
types of interacting nucleotides (A, C, G, U). The magnitude of the temperature dependence of
the stacking, $\kappa$, and the cross stacking interaction strength $\gamma$ are
set to be the same for all nucleotide types.%

In the first step of the fitting procedure, we parametrize the model to reproduce the melting temperatures of the averaged NN-model, for which we define the enthalpy and entropy contribution per base-pair step by averaging contributions of all possible Watson-Crick base-pair steps
in the NN-model. In calculating average melting temperatures of different
motifs, such as hairpins or terminal mismatches and internal mismatches, the
additional entropy and enthalpy contributions for a particular motif in the NN-model were again
averaged over all possible combinations of bases. In the averaged NN-model,
the melting temperature is hence independent of the particular sequence, but depends only on the lengths of the sequence and the particular secondary structure 
motif.

The fitting of the interaction strength parameters was done by a simulated annealing algorithm, which aims to find a set of parameters that minimizes the sum of absolute differences between the melting temperatures of a set of systems as calculated by oxRNA and as predicted by the NN-model. The algorithm uses the reweighting method outlined in the previous section on an ensemble of states generated by a VMMC simulation to calculate melting temperatures for a particular set of parameters. This algorithm for model parametrization is described in detail in Ref.~\onlinecite{oxDNA}.

First, the oxRNA model was parametrized to reproduce averaged NN-model
melting temperatures of structures with only Watson-Crick base pairs. The interaction strength
$\eta_{ij}$ was hence set to $\eta_{\rm avg}$ for all base pair types $i$ and
$j$ and $\alpha_{ij}$ was set to $\alpha_{\rm avg}$ for Watson-Crick
complementary nucleotides and $0$ otherwise. The initial values for $\alpha_{\rm avg}$, $\eta_{\rm avg}$ and $\gamma$ were first chosen by hand and then refined based on results of VMMC simulations in order to reproduce melting temperatures as predicted by the averaged NN-model of short duplexes of lengths $5$, $6$, $7$, $8$, $10$ and $12$ bp and of duplexes of lengths $5$,
$6$, $8$ bp with one overhanging nucleotide at either both 3$^\prime$ ends or both $5^\prime$
ends. 
We set $\kappa$ to be equal to $1.9756$ (in the inverse of the energy unit used by our simulation code, as defined in Supplementary Material \ref{appendix_potentials}), the
same value as was used by the previous oxDNA model.\cite{oxDNA,Ouldridge2011} We
found that leaving $\kappa$ as a free fitting parameter did not lead to a
significantly better fit to the considered sequences and motifs.

We note that for some applications, where one is more
interested in the qualitative nature of the studied system or one wants to average
over all possible sequences, it might be more useful to study the system
with a sequence independent model. We refer to such a model as the ``average-base'' model meaning that $\eta_{ij}$ are
set to $\eta_{\rm{avg}}$ for all types of bases and $\alpha_{ij}$ are set to $\alpha_{\rm{avg}}$ for all Watson-Crick
complementary base pairs (GC and AU) and $0$ otherwise.  If one is interested in sequence-specific effects, then a more complex parametrization is necessary.

We used the final values of $\eta_{\rm{avg}}$,
$\alpha_{\rm{avg}}$  as
the initial values for fitting the sequence-dependent values
$\eta_{ij}$, $\alpha_{ij}$, with $i$ and $j$ being Watson-Crick or wobble base
pairs (AU, GC, GU). 
The parameters were fitted to an ensemble that contained oligomers and hairpins
of the above mentioned sizes, with 100 randomly generated sequences (with only Watson-Crick base pairs)
for each size, and 533 further random duplexes of lengths 5 to 12 bp containing GU wobble base pairs. 
We excluded sequences with neighboring
wobble base pairs in the fitting process as these can lead to duplexes with particularly low melting
temperatures (some of the base pair steps containing wobble base pairs are
actually destabilizing at room temperature\cite{mathews1999expanded}). We found that our model was unable
to accurately fit melting temperatures of duplexes that contain neighboring
GU/UG or UG/GU wobble base pairs, probably due to the fact that we
do not account for the structural changes that these induce in the duplex. 

We note that if one included only Watson-Crick base pairs in the sequence-dependent fitting (as was the case for the parametrization of the oxDNA model\cite{oxDNA}), it would not 
be possible to distinguish between certain stacking interaction types. For instance, the contribution of AA and UU base stacking interactions always appear together in the AA/UU base pair step free-energy contribution in the NN-model. However, including wobble base pairs in the fitting ensemble provides additional information, for example the UU stacking contribution also appears in the AG/UU base-pair step. We therefore do not need to restrict the strength of stacking interaction to be the same for certain types of nucleotides, as was the case for the oxDNA model. 

Finally, we parametrized the coaxial stacking interaction potential, $V_{\rm{coaxial~st.}}$, which  captures the stacking interaction between two bases that are not
neighbors along the same strand.  Experiments have measured this interaction by a comparison of the melting of a 4-base strand with its complement, or with a hairpin with a 4-base overhang with the complementary sequence adjacent to the hairpin stem.
They found 
for both DNA and RNA that the melting temperature increases for the 4bp long strand attached to the overhang on the hairpin stem, which was attributed to the extra stabilizing interactions with the adjacent stem.\cite{walter1994sequence,
walter1994coaxial, SantaLucia2004, pyshnyi2004influence} 
The coaxial stacking
free energy has been incorporated in the NN-model by assuming that the
free-energy stabilizations in these experiments are similar in strength to the actual
base-pair steps with the same sequence. The NN-model hence uses the same free
energy contribution for a base pair coaxially stacked on a subsequent base pair
(as illustrated in Fig.~\ref{fig_potentials}) as it uses for a base
pair step in an uninterrupted duplex.  In order to parametrize these interactions for oxRNA, 
we performed melting simulations of a 5-base strand, which was able to associate with a complementary $5^{\prime}$
overhang on a longer duplex (which itself was stable). We fitted the interaction strength $\mu$ of the coaxial
stacking interaction in our model so that it would  match the prediction of the melting
temperature by the averaged NN-model.  
We note that in our model, the contributing
factors to stabilization are both the coaxial stacking interaction and the
cross-stacking interaction between the 5-base strand and the
hairpin.

\begin{table}
 \centering
 \begin{tabular}{c c c}
 \hline \hline {\bf Motif} & $ T_m - T_m({\rm NN^{avg}})$ &  $T_m [\celsius]$  \\ \hline 
  5-mer & 0.8 & 26.4 \\
  6-mer &  0.3  & 42.5   \\ 
  7-mer & 0.1  & 53.6   \\
  8-mer & -0.6  & 61.2   \\
  10-mer & -0.5 & 72.5   \\
  12-mer & -0.8 & 79.3   \\
  6-mer ($3^{\prime}$ overhangs) &  -1.1  &  49.8  \\
  6-mer ($5^{\prime}$ overhangs) &  -2.6  &  43.1  \\
  8-mer ($3^{\prime}$ overhangs) &  -0.6  &  65.6  \\
  8-mer ($5^{\prime}$ overhangs) &  -2.9  &  62.0  \\
  8-mer (terminal mismatch)   & -1.8 & 57.8 \\
 \hline
 \hline
 \end{tabular}
\caption{ The melting temperatures of a series of duplexes for the average-base parametrization of oxRNA ($T_m$) compared  to the averaged NN-model
($T_m({\rm NN^{avg}})$). The melting temperatures were calculated from VMMC simulations and are for a strand concentration of $3.36 \times 10^{-4}\,\rm{ M}$. For structures with overhangs, two single-base overhangs were present either at the $3^{\prime}$ or $5^{\prime}$ ends. The 8-mer with a terminal mismatch had a non-complementary base pair at one of the ends of the duplex. \label{table_averagedmodel}
}
\end{table}

\section{Properties of the model}
\label{sec_model_proper}
In this Section, we describe the structural properties of the model and report the thermodynamics of duplexes, hairpins and other secondary structure motifs as represented by
the model. We further study some of its mechanical properties, namely the persistence length of a duplex, the force-extension curve for duplex stretching, and the overstretching transition.

\subsection{Structure of the model}
\label{sec_structure}
As mentioned in Section \ref{sec_parametrization}, the coarse-grained interactions were selected so that the model reproduces the A-form helix that RNA duplexes have been shown to adopt at physiological
conditions.\cite{Saenger1984,neidle2010principles} 

The A-RNA structure is significantly different from  B-DNA, the prevalent duplex structure found
in DNA molecules. These differences are mainly caused by the sugars in A-RNA adopting a more twisted
conformation (C3$^\prime$ \textit{endo} pucker) as a result of the presence of
an extra OH group on the sugar. The A-RNA duplex has a reported helical twist ranging\cite{neidle2010principles} from  $32.7^{\circ}$ to $33.5^{\circ}$ per base pair,
corresponding to a pitch of $10.7$ to $11$ base pairs. The rise per base pair reported by X-ray
measurements\cite{neidle2010principles} is about $0.28\,\rm{nm}$. The bases are
displaced from the helical axis, i.e.\ the helical axis does not pass through the
base pair mid-points as it is approximately the case for the B-DNA helix. Finally, the bases are
not perpendicular to the helical axis, but are inclined at an angle of about $15.5^{\circ}$.
Although the width of A-RNA is reported to be about $2.1\,\rm{nm}$
from X-ray crystal structures,\cite{neidle1999oxford} Reference
\onlinecite{kebbekus1995persistence} uses an effective hydrodynamic diameter of $2.8\,
\rm{nm}$ for the structure.
The A-RNA helix has a narrow major groove ($0.47\,\rm{nm}$) and a wide minor groove
($1.08\,\rm{nm}$). 

To characterize the structure of the oxRNA duplex, we simulate a 13-base-pair duplex at
$25\,^{\circ}{\rm C}$ using Monte Carlo simulation. We generated $30\,000$ decorrelated configurations that were analyzed in the following manner. 
The helical axis was fitted for each saved configuration, as described in Supplementary Material \ref{app_axis_fit}. 
The rise per base pair
was measured as the distance between the projections of the midpoints of base pairs onto the helical
axis. The length scale in the oxRNA model is defined so that the rise per base pair is $0.28\,\rm{nm}$. The twist per base
pair was measured as the angle between the projections of the vectors connecting bases in the base pairs onto the
plane perpendicular to the helical axis. The mean turn per base pair in the model is
$33.0^{\circ}$, corresponding to a pitch of $10.9$ base pairs. The inclination, measured as
the mean angle between the vector pointing from the center of mass of a nucleotide to its base and the plane perpendicular to the helical axis, is $16.1^{\circ}$. 

The width of the helix is measured as twice the distance of the backbone from
the axis, and includes the excluded volume interaction radius of the
backbone site. The helix width in oxRNA is $2.5\,\rm{nm}$.  The major and minor
grooves in oxRNA are $0.48\,\rm{nm}$ and $1.07\,\rm{nm}$, respectively, where we measured the groove distances in a manner analogous to a method employed by the
Curves+ software\cite{Lavery2009} for analyzing atomistic structures of DNA and
RNA. For a selected nucleotide, we measured distances between its backbone site and points on a curve that was linearly interpolated through the backbone sites of the nucleotides on the opposite strand. The distances measured along the curve have two minima, one for each groove. 
The excluded volume interaction radius for each backbone site was subtracted from these measured distances. 

\subsection{Thermodynamics of the model}
\label{sec_thermo}
In this section, we examine the thermodynamics of duplexes, hairpins, bulges,
and internal and terminal mismatches as represented by oxRNA. We compare the
melting temperatures as predicted by oxRNA with the melting temperatures
calculated from the NN-model (denoted as $T_m({\rm NN})$) for different sequences
and different secondary structure motifs. To calculate the melting
temperatures, we used the reweighting method, described in Section \ref{sec_rewei}, with the states that were generated from VMMC simulations of melting for the average-base parametrization of oxRNA.

\subsubsection{Duplex and hairpin melting}

A comparison of the melting temperatures of the average-base parametrization of oxRNA
with the thermodynamics of the averaged NN-model for structures involving only Watson-Crick base pairs is
shown in Table \ref{table_averagedmodel}.  For this averaged model, the differences are roughly on the order of the accuracy of the NN-model itself.

\begin{figure}
\includegraphics[width=0.5\textwidth]{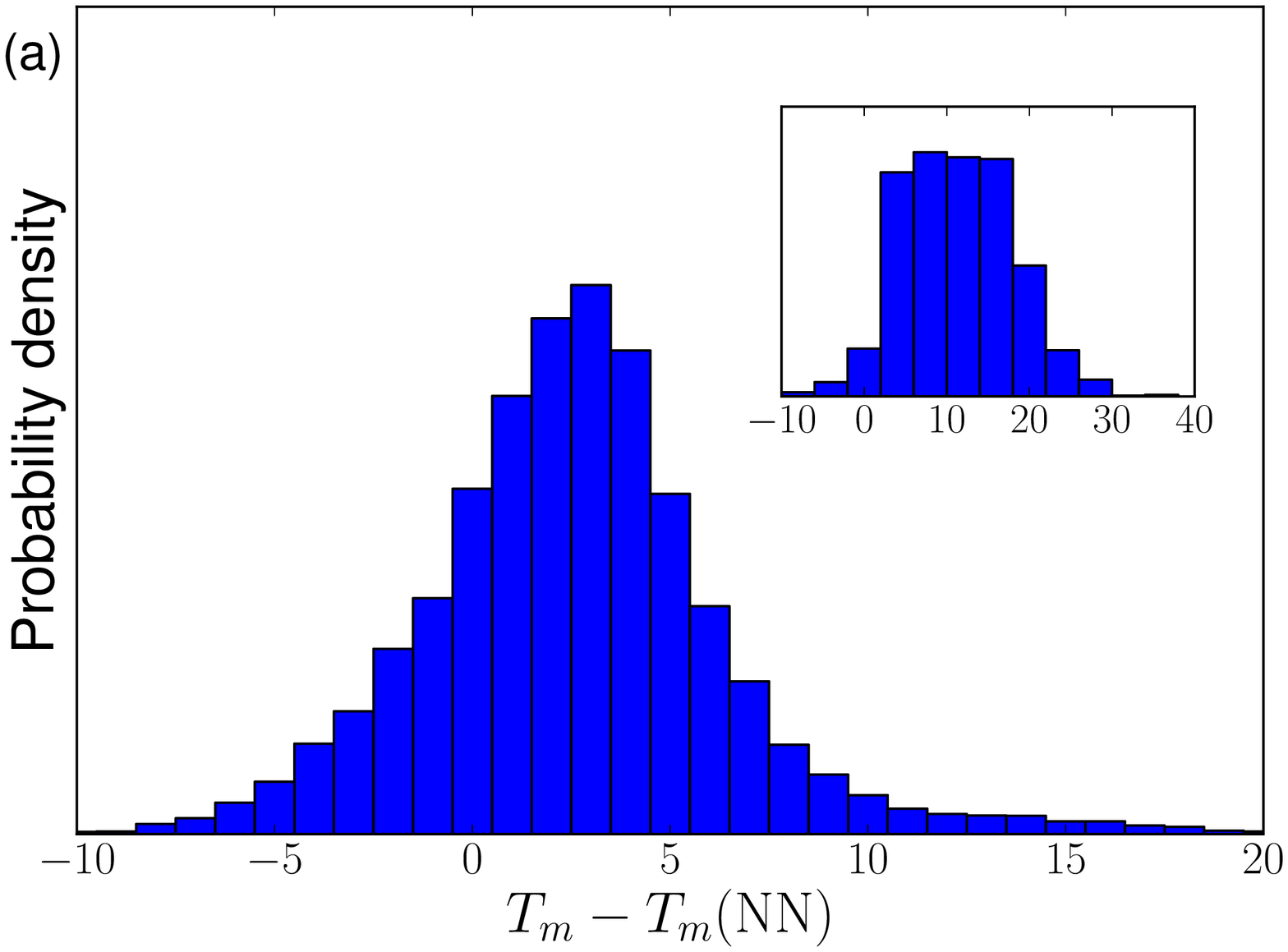}
\includegraphics[width=0.5\textwidth]{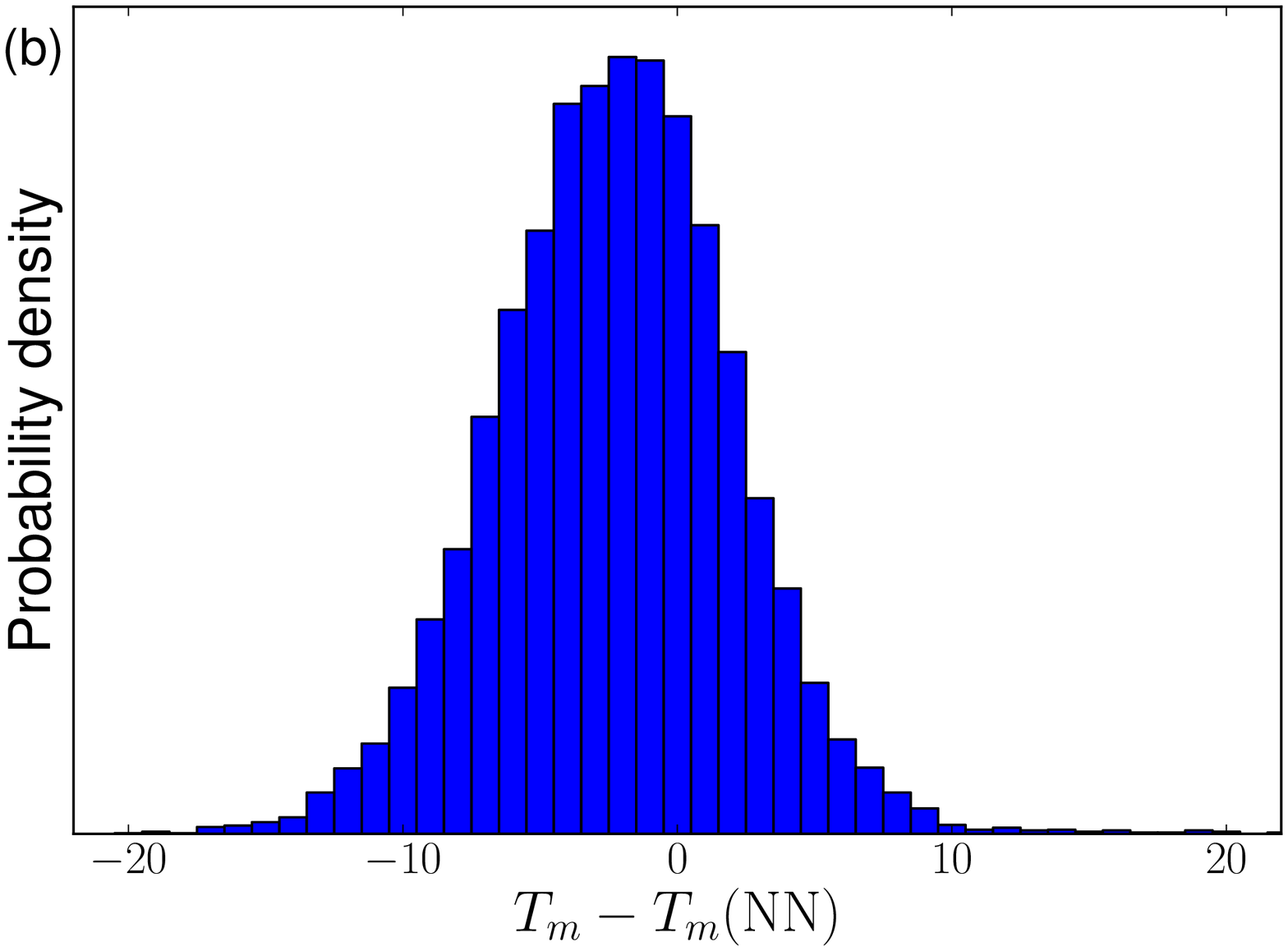}
\centering
\caption{(a) The histogram of differences between melting temperatures as predicted by the oxRNA
model ($T_m$) and by the NN-model ($T_m({\rm NN})$) for a set of  $20\,255$  randomly
generated RNA duplexes of lengths 6, 7, 8, 10 and 12 with Watson-Crick and
wobble base pairing. The main plot shows a histogram of values of $T_m$ -
$T_m({\rm NN})$ for duplexes that do not include GU/UG or UG/GU base pair steps.
The inset shows a histogram of values of $T_m$ - $T_m({\rm NN})$ for 1439 randomly
generated sequences that contained at least one GU/UG or UG/GU base pair steps. 
(b) The histogram of differences between melting temperatures as
predicted by the oxRNA model and by the NN-model for a
set of $12\,000$ randomly generated hairpins with stems
of lengths 6, 8, and 10 and loops with lengths of 5, 6, 7, 8, 10 and 15, where the stems only contain Watson-Crick base pairs}
\label{fig_duplexhairpinhist}
\end{figure}

To test the sequence-dependent parametrization of the hydrogen-bonding and stacking strength of the
interactions, we calculated the melting temperatures for randomly generated
ensembles of RNA duplexes, different from the ensemble used for parametrization. 
A histogram of the differences in the melting
temperature predicted by the sequence-dependent version of oxRNA ($T_m$) and those calculated from
the nearest neighbor model ($T_m({\rm NN})$) is shown in Figure
\ref{fig_duplexhairpinhist}(a). The main histogram is for duplexes 
with both Watson-Crick and wobble base pairing, but not containing GU/UG or UG/GU base pair steps. For
convenience, the generated ensembles of sequences also do not include any self-complementary
sequences because the calculation of their melting temperatures requires
a different finite size correction.\cite{ouldridge2012inferring,Ouldridge_bulk_2010} The average difference
in melting temperatures is $2.1\celsius$, with an average
absolute deviation of $3.4\celsius$. The histogram in the inset of Fig.~\ref{fig_duplexhairpinhist}(a) is for
sequences containing at least one GU/UG or UG/GU base pair step.
The average
difference in melting temperatures for the ensemble is $9.4\celsius$ and the
absolute average deviation is $9.7\celsius$.
We note that in
the NN-model, the free-energy contribution of these base pair steps is positive
at $37\celsius$, meaning that they actually destabilize the
duplex. However, in the oxRNA model, the cross-stacking, stacking and hydrogen-bonding interactions are always stabilizing interactions and
the interaction strength of two hydrogen-bonded nucleotides does not depend on the identity of their respective neighbors on the strand.
Our coarse-grained model hence cannot capture the free-energy contributions of GU/UG and UG/GU base pair steps.
One could imagine adding multi-body interactions, but for the sake of computational efficiency and maintaining the consistency of our coarse-graining methodology, we do not do so in this study. Another option might be to introduce a structural perturbation of the helix caused by the GU base pairs.

The histogram in Fig.~\ref{fig_duplexhairpinhist}(b) shows the
difference between melting temperatures calculated by oxRNA and those predicted
by the NN-model for an ensemble of 
randomly generated hairpins.
The average melting temperature difference was
$-2.8\celsius$ with the average absolute deviation being $4.1\celsius$.

The transition widths for duplex and hairpin formation were calculated
for the averaged model as the difference between the temperatures at which the
yield is 0.8 and 0.2, respectively. This quantity is important because the widths of the transition determine the change of the duplex melting temperature with concentration. It can be shown\cite{Ouldridge2011} that the derivative of the melting temperature as a function of strand concentration is proportional
to the width of the transition.
The melting simulations of oxRNA with the average-base parametrization were compared with the width
predicted by the averaged NN-model. For the duplexes of lengths 6, 8, 10, and 12 the width was on average underestimated by $0.9\celsius$, but was overestimated for a 5-mer by $0.5\celsius$. 
The width of the transition for the averaged NN-model decreases from $20.5\celsius$ for a 5-mer to $9.2\celsius$ for a 12-mer.
For a set of hairpins with stems of length 6, 8, and 10 bp and with loops of lengths 5, 6, 7, 8, 10 and 15, the width of the melting transition was on average underestimated by  
$1.5\celsius$. The width of the hairpin transition decreases from about $12\celsius$ for stems of length 6 to approximately $8\celsius$ for stems of lengths 10 in the averaged NN-model.
However, the trend of increasing width with decreasing size is always captured by the oxRNA model.

Finally, we note that we could have parametrized the sequence-dependent model only to duplex melting temperatures, as for oxDNA, which would then have led to a larger underestimate of hairpin melting temperatures, presumably because our representation of the strand is too simple to exactly capture the entropy and enthalpy of the loop formation. We hence chose to parametrize to the 
ensemble of duplexes and hairpins because hairpins are a prominent secondary structure motif in RNA.

\subsubsection{Thermodynamics of secondary structure motifs}
Given that our aim is to design a model that goes beyond describing
hybridization of simple duplexes, it is important to assess how well it is able
to reproduce the thermodynamic variation induced by common secondary structure
motifs such as bulges or mismatches. To this end, we have studied the
melting temperature changes induced by internal mismatches, terminal mismatches
and bulges of different lengths.

To assess the effects of bulges, we consider a systems of two strands, one with
10 bases and the other with 10
complementary bases and extra bases that create a bulge motif. We
considered bulges of lengths 1, 2, 3 and 6, positioned in the center. For each sequence
considered we calculated the melting temperatures by reweighting a set of 6000
states that were sampled from a melting simulation using the average-base 
parametrization. For each bulge length, we considered 1000 randomly generated
sequences with Watson-Crick base pairing in the complementary part. 

We further
evaluated the melting temperatures for randomly generated 10-mers which
contained 1, 2 or 3 consecutive mismatches (and therefore had 9, 8 or 7
complementary Watson-Crick base pairs). The average difference and absolute
average deviation for the considered bulges and mismatches are shown in Table \ref{table_ssmotifs}.
The melting temperatures of duplexes with bulges are underestimated by a few degrees. However, the model presently significantly overestimates the stability of 
internal mismatches by the order of $10\celsius$ or more.  
Even though the mismatching base pairs do not gain stabilization from hydrogen-bonding interactions (which are zero for bases that are not complementary), they still have favorable
cross-stacking and stacking interactions, which have their minimum energy configuration in an A-helical configuration, which the oxRNA model can still form with the mismatches presents. We further note that our model represents each nucleotide by the same rigid body structure, so the mismatching base pairs do not cause any distortion to the duplex structure in our model. Improving the model to more accurately represent the secondary structures with mismatching nucleotides will be the subject of future work.

\begin{table}
 \centering
 \begin{tabular}{c c c c c}
 \hline \hline {\bf Motif} & $\left\langle\Delta T_m \right\rangle$ &  $\left\langle  \left| \Delta T_m \right| \right\rangle$ & $ \Delta T_m^{{\rm duplex}}(\rm{NN})$ & $\Delta T_m^{{\rm duplex}}$ \\
 \hline 
 Bulge (1 b) & -3.6 & 4.3        & -13.0     &  -19.5   \\
 Bulge (2 b) & -2.0 & 4.3        & -17.5     &  -22.4 \\
 Bulge (3 b) & -4.3 & 5.9        & -19.7     &  -27.0 \\
 Bulge (6 b) & -0.8 & 5.2        & -25.4     &  -29.1 \\
 Internal mis. (1 b) & 10.2  & 10.2 & -18.0  &  -10.8 \\
 Internal mis. (2 b) & 8.9   &  9.5 & -27.4  &  -21.2 \\
 Internal mis. (3 b) & 16.7  &  16.8 & -45.1 &  -31.5  \\
 \hline
 \hline
 \end{tabular}
\caption{Melting temperatures for bulge and internal mismatch motifs in a 10-mer. $\left\langle\Delta T_m \right\rangle = \left\langle T_m - T_m({\rm NN}) \right\rangle$ is the average difference between the melting temperature of the oxRNA model ($T_m$) and the melting temperature as predicted by the NN-model ($T_m({\rm NN})$).  
$\left\langle \left| \Delta  T_m \right| \right\rangle = \left\langle \left| T_m - T_m({\rm NN}) \right| \right\rangle$ is the average absolute difference in melting temperatures.
 $ \Delta T_m^{{\rm duplex}}(\rm{NN})$ and $\Delta T_m^{{\rm duplex}}$ are the average differences in melting temperature between the sequences with a secondary structure motif and a duplex with the same sequence but with no bulge or internal mismatch as predicted by the NN-model and oxRNA respectively.
Each of the motifs considered is destabilizing, resulting in a decrease of the melting temperature.
The averages were taken over an ensemble of
randomly generated sequences (1000 for each motif) that had 10 complementary Watson-Crick base pairs
for the bulges, and 9, 8, and 7 complementary base pairs for internal mismatches
of size 1, 2 and 3 bases, respectively. The bulges that we consider were of the size 1, 2, 3 and 6 bases.
 All the melting temperature calculations were
calculated for an equal strand concentration of $3.36 \times 10^{-4}\,\rm{ M}$. \label{table_ssmotifs}}
\end{table}

\subsection{Mechanical properties of the model}

\subsubsection{Persistence length}

The persistence length $L_p$  
of dsRNA molecule measured in experiments is reported to be between $58\,{\rm nm}$ to
$80\,{\rm nm}$,\cite{abels2005single,hagerman1997flexibility,herrero2012mechanical} corresponding to
$206$--$286$ bp (assuming $0.28\,$nm rise per base pair). The first studies of the persistence length of dsRNA used electron micrographic, gel-based and hydrodynamic measurements  (reviewed in Ref.~\onlinecite{hagerman1997flexibility}) and reported the persistence length to be between 70 to 100\,nm, in salt conditions ranging from $6\,$mM [Mg$^{2+}$] and $0.01\,$M to $0.5\,$M [Na$^{+}$]. 
A more recent single-molecule experimental study\cite{abels2005single} in $0.01\,$M [Na$^{+}$] and $0.01\,$M [K$^{+}$] buffer measured the persistence length by analyzing force-extension curves in magnetic tweezers experiments as well as by analyzing atomic force microscopy images of the RNA duplexes. The two methods yielded consistent values with the measured persistence length estimated as $63.8$ and $62.2\,{\rm nm}$ respectively, corresponding to 227 and 222 bp. Finally, a recent single molecule force-extension study\cite{herrero2012mechanical} of dsDNA and dsRNA at salt concentrations ranging from $0.15\,$M to $0.5\,$M [Na$^{+}$] found the dsRNA persistence length to decrease from $67.7$ to $57.7\,{\rm nm}$ with increasing salt concentration, and the extrapolation of measured persistence lengths to higher salt concentrations approaches asymptotically $48\,{\rm nm}$ (171 bp). 

\begin{figure}
\includegraphics[width=0.5\textwidth]{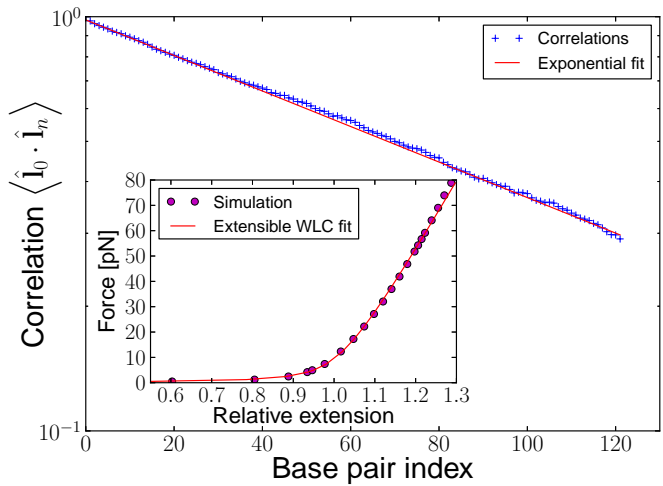}
\centering
\caption{Semi-logarithmic plot of the correlation function for the direction of the local helix axis along the duplex as defined in Eq.~\ref{eq_correl} and an exponential fit to the data. The inset shows the force-extension curve of a 81-bp section of a 99-bp duplex. The extension is normalized with respect to the contour length of the 81-bp duplex (with a rise of $2.8\,{\rm nm}$ per bp) and a fit to the data using the
extensible wormlike chain model defined in Eq.~\ref{wormlike_extensible} is also plotted.}
\label{fig_persis}
\end{figure}

To measure the persistence length in our model, we simulated an 142-bp long double-stranded
RNA with the average-base oxRNA model, and measured the correlations in the
orientation of a local helical axis along the strand. The local axis vector
${\hat{\mathbf{l}}_i}$ is fitted through the $i$-th base pair and its nearest neighbors, using the approach described in Supplementary Material \ref{app_axis_fit}, but considering only the 
nearest neighbors. We found the results to be robust even when 2 or 3 next nearest neighbors were included in the construction of the local axis.
To account for edge effects, a section of ten base pairs
at each end of the duplex was ignored when accumulating averages. For an
infinitely long, semi-flexible polymer in which the correlations decay
exponentially with separation along the strand, the persistence length $L_p$ can
be obtained from\cite{edwards1986theory} 
\begin{equation}
\label{eq_correl}
 \left\langle \hat{\mathbf{l}}_0 \cdot \hat{\mathbf{l}}_n \right\rangle = \exp \left( - \frac {n \left\langle r \right\rangle}{L_{p}}\right)
\end{equation}
where
$\langle r \rangle$ is the
rise per base pair. This correlation function is
shown in Fig.~\ref{fig_persis} along with the exponential fit 
from which we estimated the persistence length of our model to be about 101
base pairs, somewhat lower than the $171$ bp persistence length expected at this salt concentration.
Our model's persistence length is hence smaller than the experimentally measured values, but still within a factor of 2.
OxRNA hence captures the correct order of magnitude for the persistence length and as long as one studies structures whose
duplex segments are smaller than the persistence length of the model,
it should provide a physically reasonable description. 

We note that the persistence length is quite hard to correctly reproduce. We expect this issue to hold for other coarse-grained models as well. To our knowledge, the persistence length has not been measured yet in these models. 

\subsubsection{Force-extension properties}
As a further test of the mechanical properties of the model, we measured the
extension of a 99-bp RNA duplex as a function of applied force for the average-base model. We
used a constant force acting on the center of mass of the first and last
nucleotides in one of the two strands in the duplex. We focus on the average
extension between the 11$^{\rm th}$ and 91$^{\rm st}$ nucleotide on this strand
in order to avoid end effects, such as from fraying of base pairs, in our measurements.

We compare our force-extension data with an extensible worm-like chain
model,\cite{Odijk1995} which provides the following expression for the
projection of the end-to-end distance ${\mathbf R}$ along the direction of the
force $\hat{{\mathbf z}}$:
\begin{equation}
\label{wormlike_extensible}
 \left\langle \mathbf{R} \cdot \hat{\mathbf{z}} \right\rangle = L_0 \left( 1 + \frac{F}{K} - \frac{k_{\rm B} T}{2 F L_0} \left(1 +  A  \coth A   \right)   \right)  
\end{equation}
where
\begin{equation*}
  A = \sqrt{\frac{F L_0^2}{L_p k_{\rm B} T}},
\end{equation*}
$K$ is the extension modulus and $L_0$ is the contour length.
This expression comes from an expansion around the fully extended state, and
thus it is expected to be valid at forces high enough for the polymer to be 
almost fully extended.

It was shown experimentally\cite{herrero2012mechanical} that this extensible worm-like chain model describes
the behavior of dsRNA prior to the overstretching transition. In particular, at
$0.5\,\rm{M}$ \sodium, the extensible worm-like chain model fit to the
experimentally measured force-extension curve
yielded $L_p = 57.7\,{\rm nm}$ and $K=615$\,pN.

The force-extension curve for oxRNA is shown in the inset of 
Fig.~\ref{fig_persis}. We used data from forces between $2.4\,{\rm pN}$ and $69\,{\rm pN}$ for our fitting and allowed $L_0$, $K$ and $L_p$ to be free parameters.
From the fit we obtained $L_0 = 23.4\,{\rm nm}$ (84 bp), $K = 296\,$pN and $L_p=26.0\,$nm
(93 bp). We note, however, that Eq.~\ref{wormlike_extensible} is not a
robust fit for our model: changing the fitting interval and thus including or 
excluding points at either high or low forces significantly changes the resulting values of the fitting parameters, even though the residual error in the fit remains small.
The persistence length, for instance, can change by more than a factor
of two. In the interval we selected for the fit, the $L_p$ obtained 
approximately corresponds to that obtained from the correlation function in Fig.~\ref{fig_persis}, but given the sensitivity of the fit (which was not observed for the oxDNA force-extension curve\cite{Ouldridge2011}), the errors on these extracted parameter values should be taken to be quite large.
Another issue to keep in mind is that the inclination angle is also affected by the force.  At 10\,pN, this is roughly a 1 to 2 degree change, but close to the point where the chain starts to significantly overstretch (as discussed in the next section), the inclination has disappeared, and the bases are almost perpendicular to the axis. It is likely that this deformation is not entirely physical. However, the physical structure of stretched RNA is not experimentally known. In DNA, the structure of the extended state is a very active topic of research.

\subsubsection{Overstretching}

\begin{figure}
\includegraphics[width=0.5\textwidth]{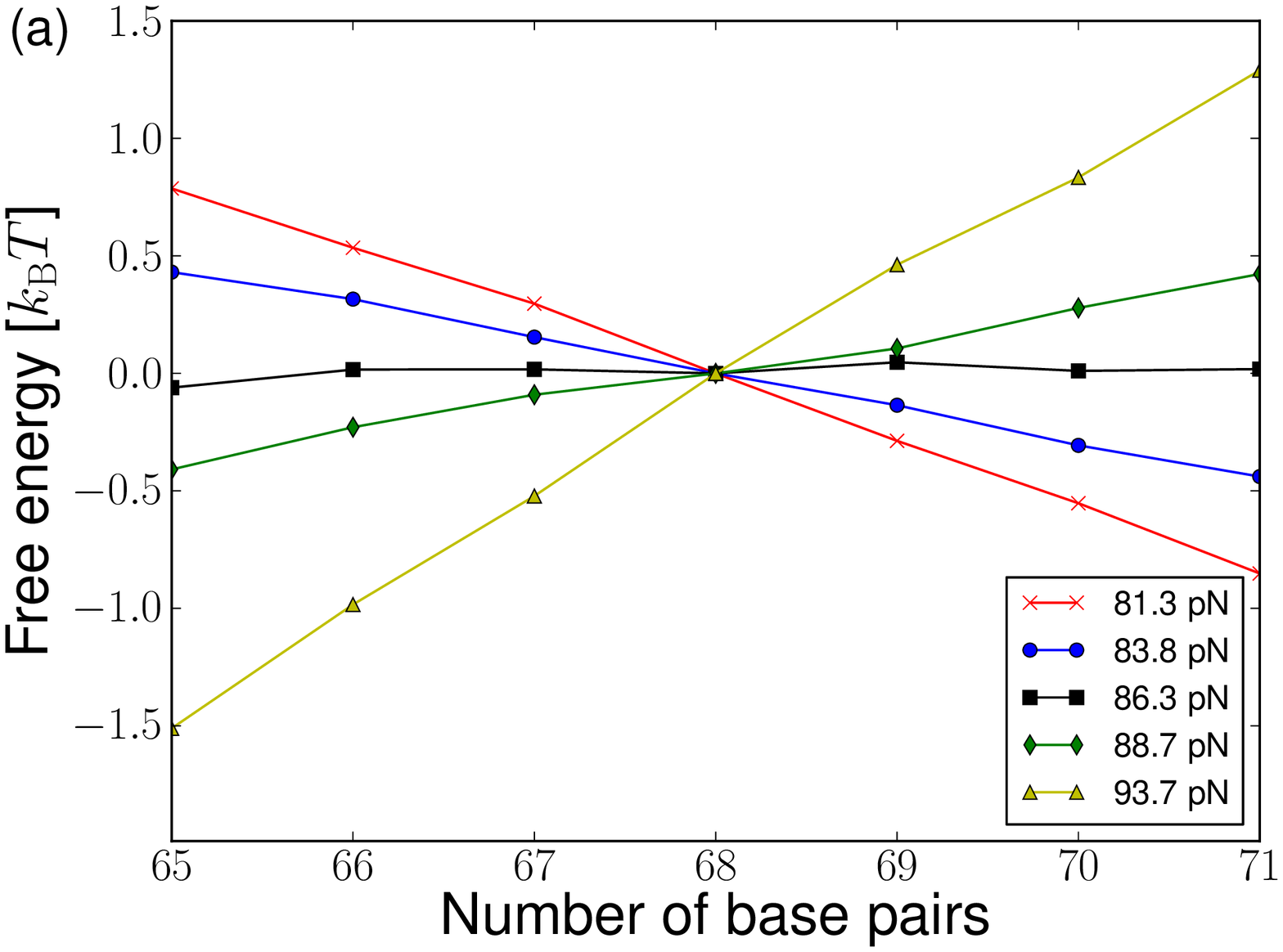}\\
\includegraphics[width=0.5\textwidth]{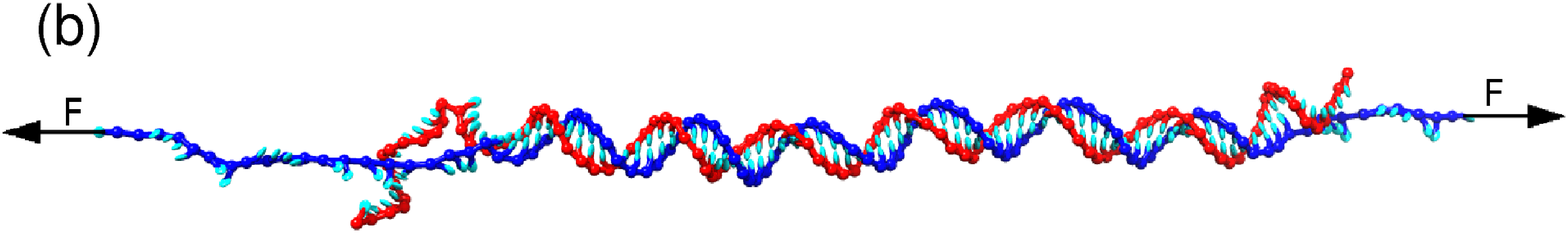}
\centering
\caption{(a) Free energy as a function of the number of base pairs in the duplex for different forces, where we set the free energy to be $0$ for 68 bp for each force considered. At the overstretching force, the slope of the free-energy profile is 0. (b) Snapshot from a VMMC simulation at $F = 86.3\,{\rm pN}$, showing unpeeling from the ends.}
\label{fig_overs}
\end{figure}

Both DNA and RNA duplexes are known to undergo an overstretching transition at high enough force.
Recent experiments \cite{herrero2012mechanical} for different salt concentrations find $63.6\,(2.0)\,{\rm pN}$ for RNA overstretching at $0.15 \,{\rm M}$ \sodium \ up to $65.9\, (3.3)\, {\rm pN}$ at $0.5\,{\rm M}$ \sodium. 
Following the approach taken in the study of DNA overstretching with the oxDNA model,\cite{romano2013coarse} we used the average-base oxRNA model to run VMMC simulations of a 99-bp RNA duplex with equal and opposite forces applied to the first and last nucleotide of one strand. 
In our simulations, only native base pairs were allowed to form to aid equilibration, i.e.\ no misbonds in the duplexes or intrastrand base pairs in the unpeeled strand.
The simulations were started from a partially unpeeled state to sample states which have between 65 and 71 bp. The obtained free-energy profiles as a function of the number of base pairs are shown in Fig.~\ref{fig_overs}(a).
As the force increases, the slope of the free energy profiles changes from negative (states with more base pairs are favored) to positive (it is favorable for duplexes to unpeel).
By estimating the force at which the slope becomes zero, we obtained $86.2\,{\rm pN}$ as the overstretching force. We note that our model was parametrized for 1\,M [Na$^{+}$], whereas the overstretching experiment was done at $0.5\,$M. Furthermore, by not allowing any formation of secondary structure in the unpeeled strands, we overestimate the overstretching force in the model, because these intramolecular base pairs stabilize the unpeeled state. For the oxDNA model, it was shown that allowing secondary structure 
decreases the overstretching force by about $3\,{\rm pN}$.\cite{romano2013coarse}  We would expect the stabilization to be slightly higher for RNA, as RNA base pairs are more stable.  
Our model hence overestimates the value of the overstretching force by about $16$-$20\,{\rm pN}$. 
This overestimation is partly due to the higher extensibility of the duplex in oxRNA, which is aided by the loss of inclination in the duplex when higher forces are applied, as we already discussed in the previous section.   

\section{Example Applications}
\label{sec_examples}

\subsection{Pseudoknots}

\begin{figure}
\includegraphics[width=0.35\textwidth]{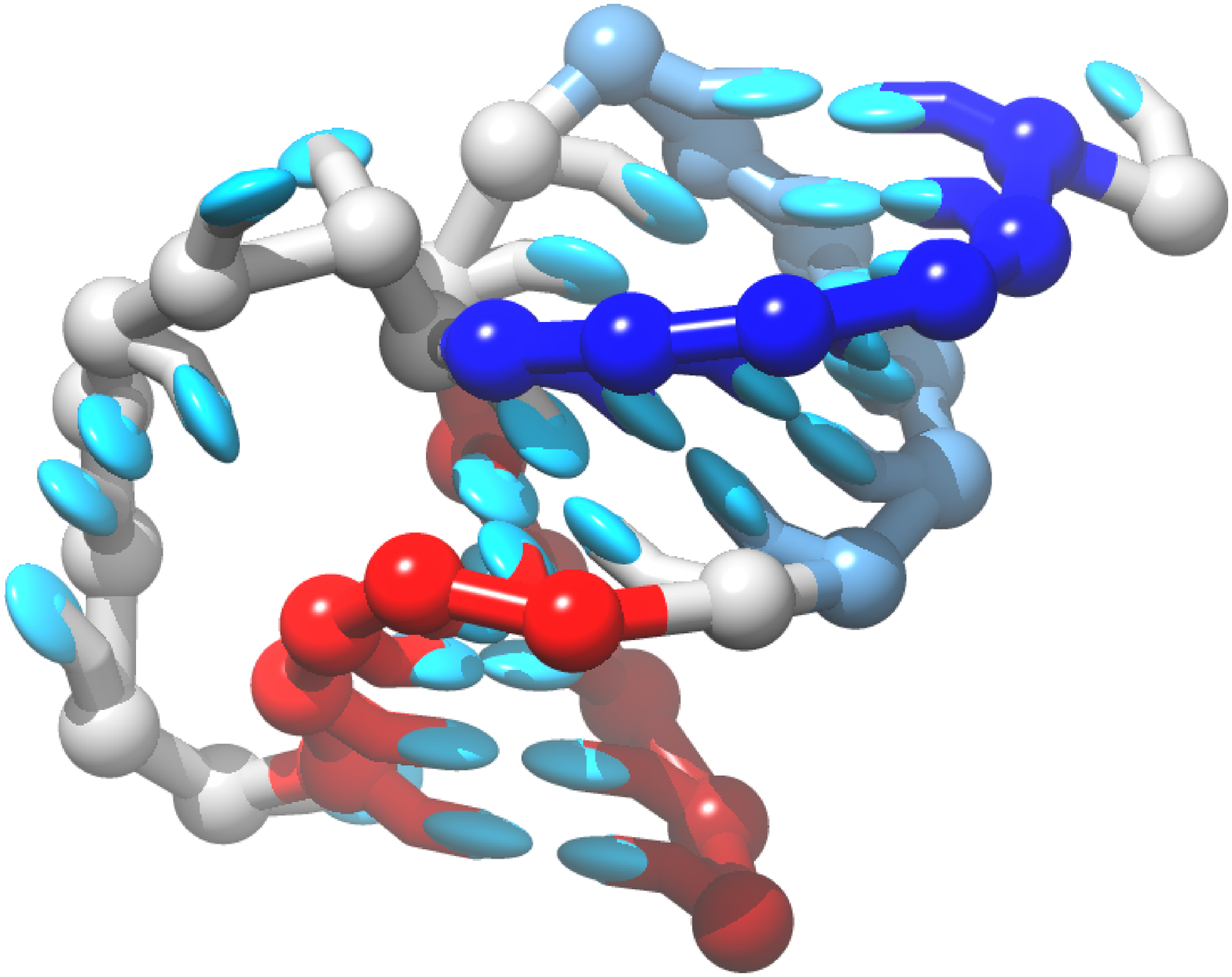}
\includegraphics[width=0.1\textwidth,height=0.3\textwidth]{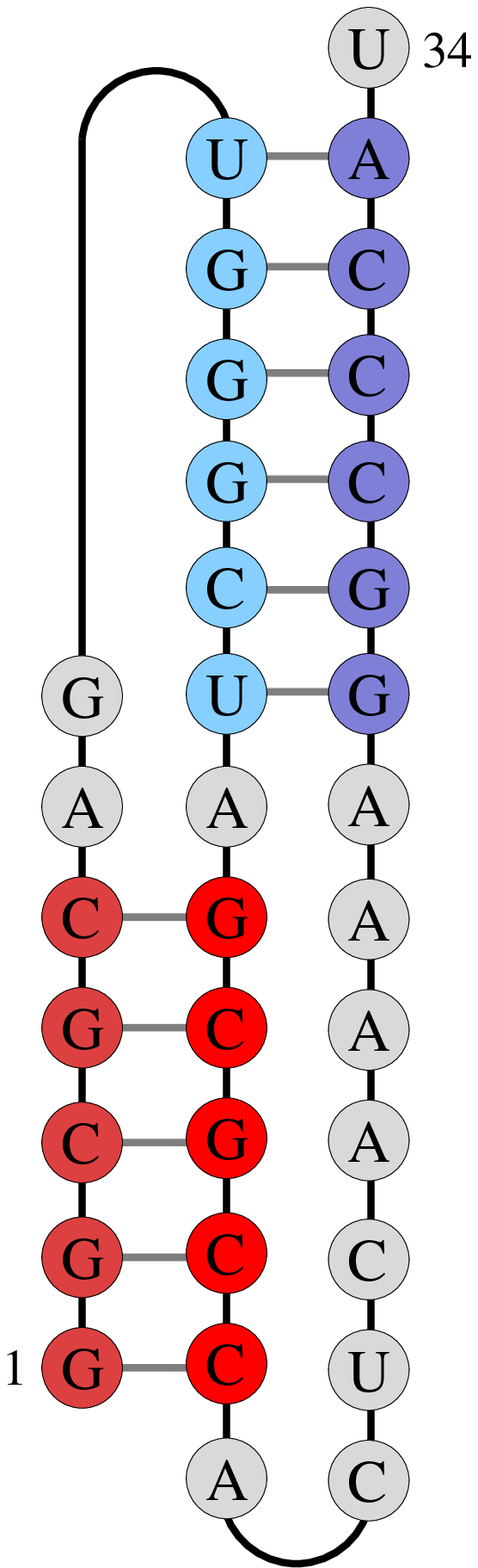}

\centering 
\caption{A snapshot of the MMTV pseudoknot as represented by oxRNA. Stem 1 (shown in blue) has 6 base pairs whereas stem 2 (shown in red) has 5 base pairs.
A schematic representation of the secondary structure with the
sequence is shown on the right.}
\label{fig_pseudo}
\end{figure}

Pseudoknots are a common structural motif in RNA. 
If a strand is represented as a circle and base pair contacts are represented as chords, then its structure is pseudoknotted if the chords cross.
Most secondary structure prediction tools do not include pseudoknotted structures in
their computations and thus cannot be used to study systems where they are
relevant, although some progress has been made in developing efficient algorithms for this task.\cite{rivas1999dynamic,bon2011tt2ne} 

Since oxRNA provides an explicitly three-dimensional representation of the RNA strands, it
can be used to simulate the folding and thermodynamics of RNA structures that
involve pseudoknots. In this section, we use our model to study the well known MMTV
pseudoknot.\cite{theimer2000contribution} The sequence and the 
three-dimensional representation of the MMTV pseudoknot by oxRNA are shown
in Fig.~\ref{fig_pseudo}. The MMTV pseudoknot's thermodynamic properties were previously studied in
experiment,\cite{theimer2000contribution} as well as with another coarse-grained RNA model.\cite{denesyuk2013coarse}
Moreover, the MMTV pseudoknot's structure has also been investigated by NMR experiments\cite{shen1995structure} and 
two stems were identified in the folded structure: stem 1 with 6 base pairs and stem 2 with 5 base pairs, as schematically shown in Figure~\ref{fig_pseudo}.

\begin{figure}
\includegraphics[width=0.5\textwidth]{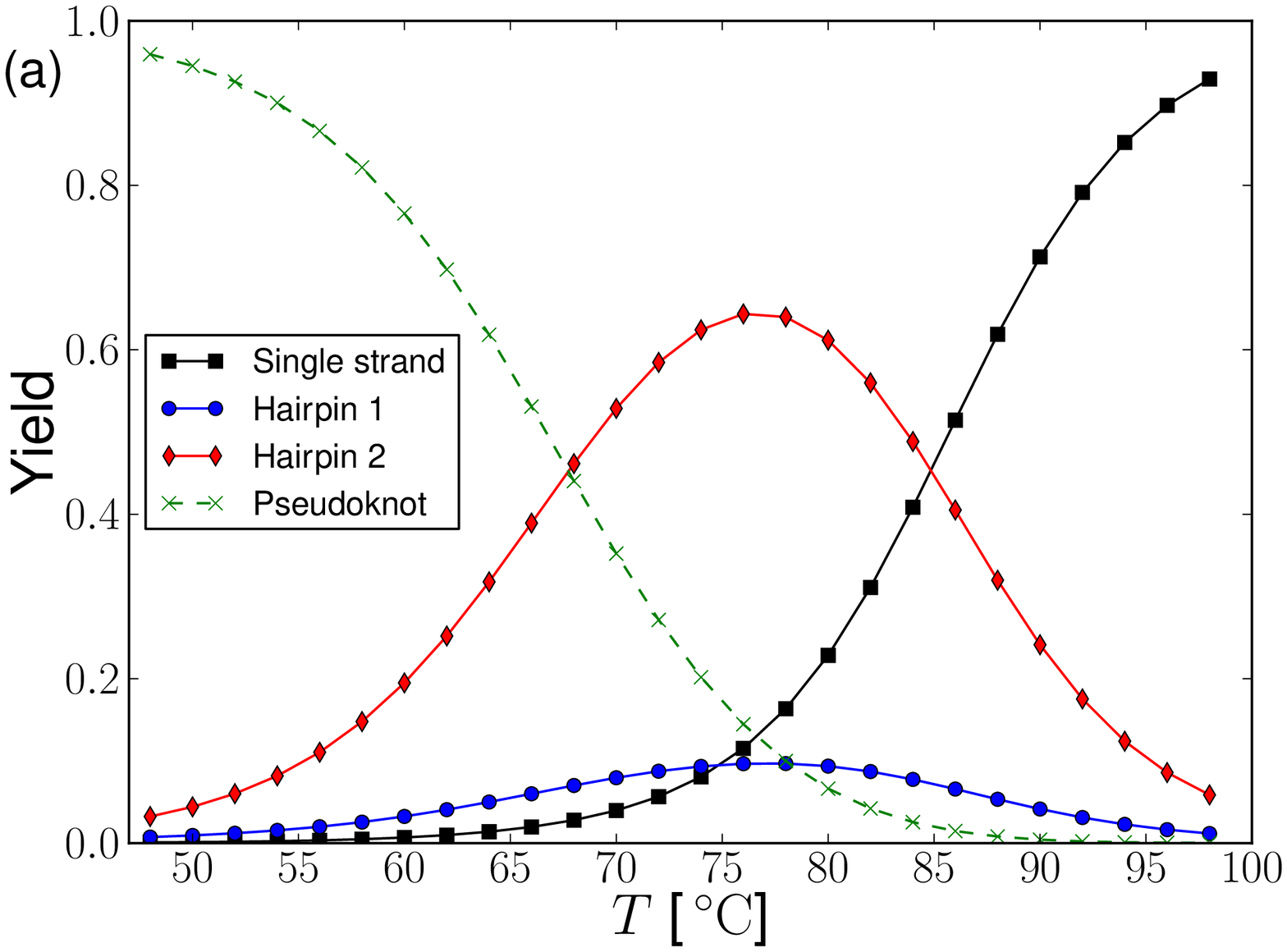}
\includegraphics[width=0.5\textwidth]{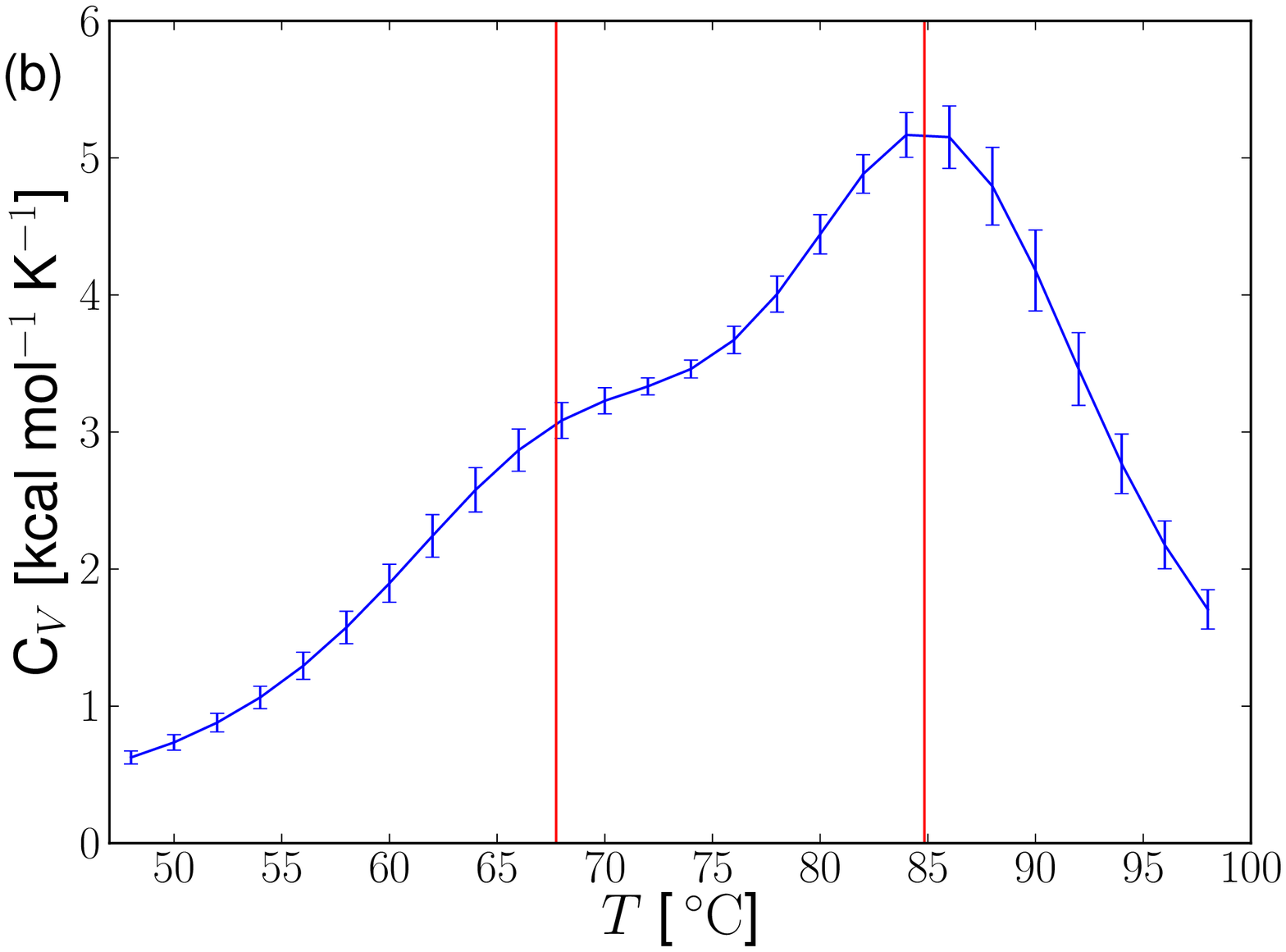}
\centering
\caption{
(a) Equilibrium yields and (b) $C_{\rm V}$ as a function of temperature for the MMTV pseudoknot. In (a) the pseudoknot and hairpins are defined as having at least 1 native 
base pair in the relevant stems, whereas the unstructured single-stranded state has no native base pairs. 
In (b) the error bars are the standard deviations derived from 5 independent simulations. The red vertical lines indicate the temperatures at which we observe equal yields of 
pseudoknot and hairpin 2 
($67.7\celsius$) and hairpin 2 and the unstructured single strand ($84.8\celsius$).}
\label{fig_pseudo_yield_cv}
\end{figure}

\begin{figure}
\includegraphics[width=0.5\textwidth]{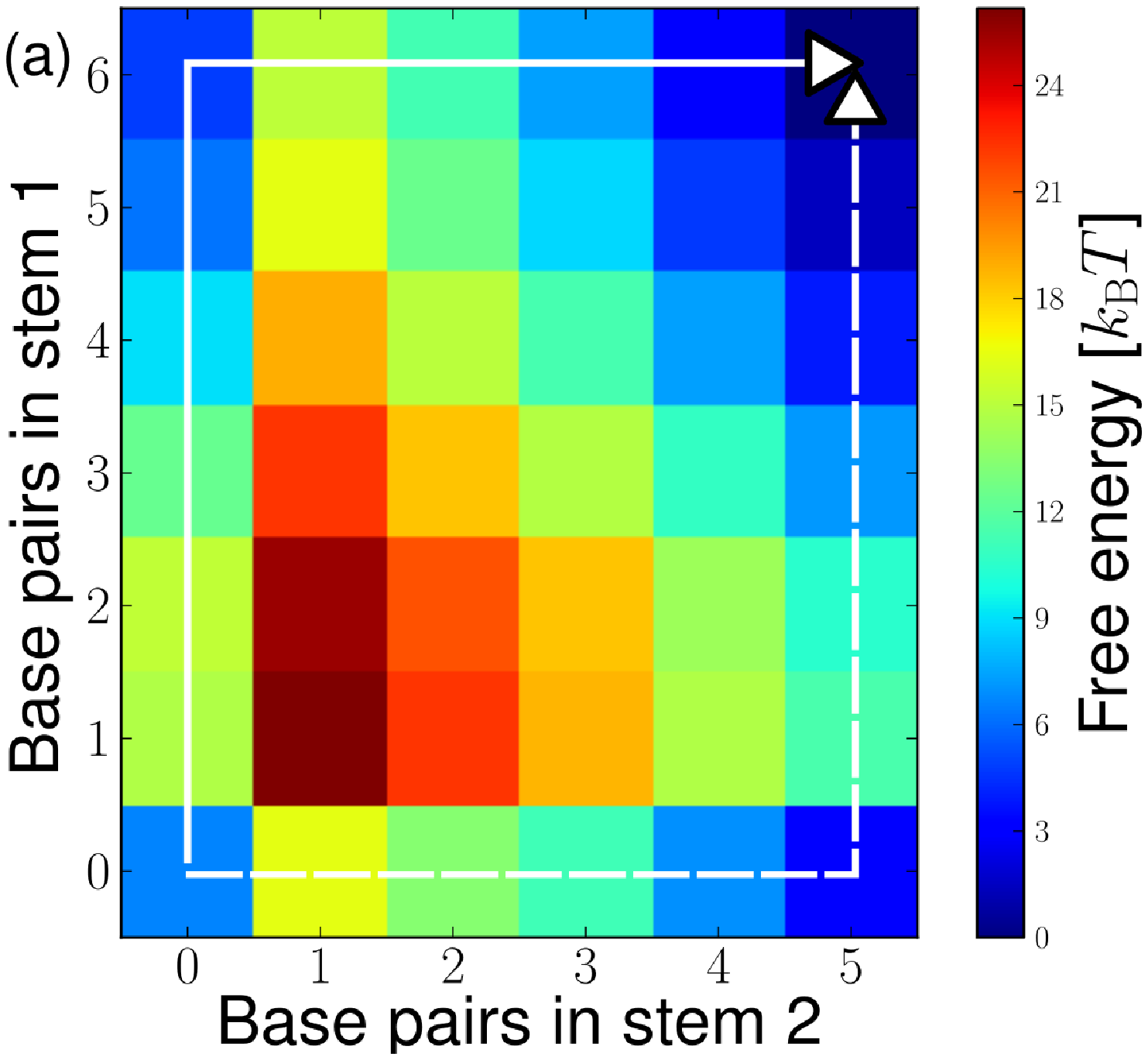}\\
\includegraphics[width=0.5\textwidth]{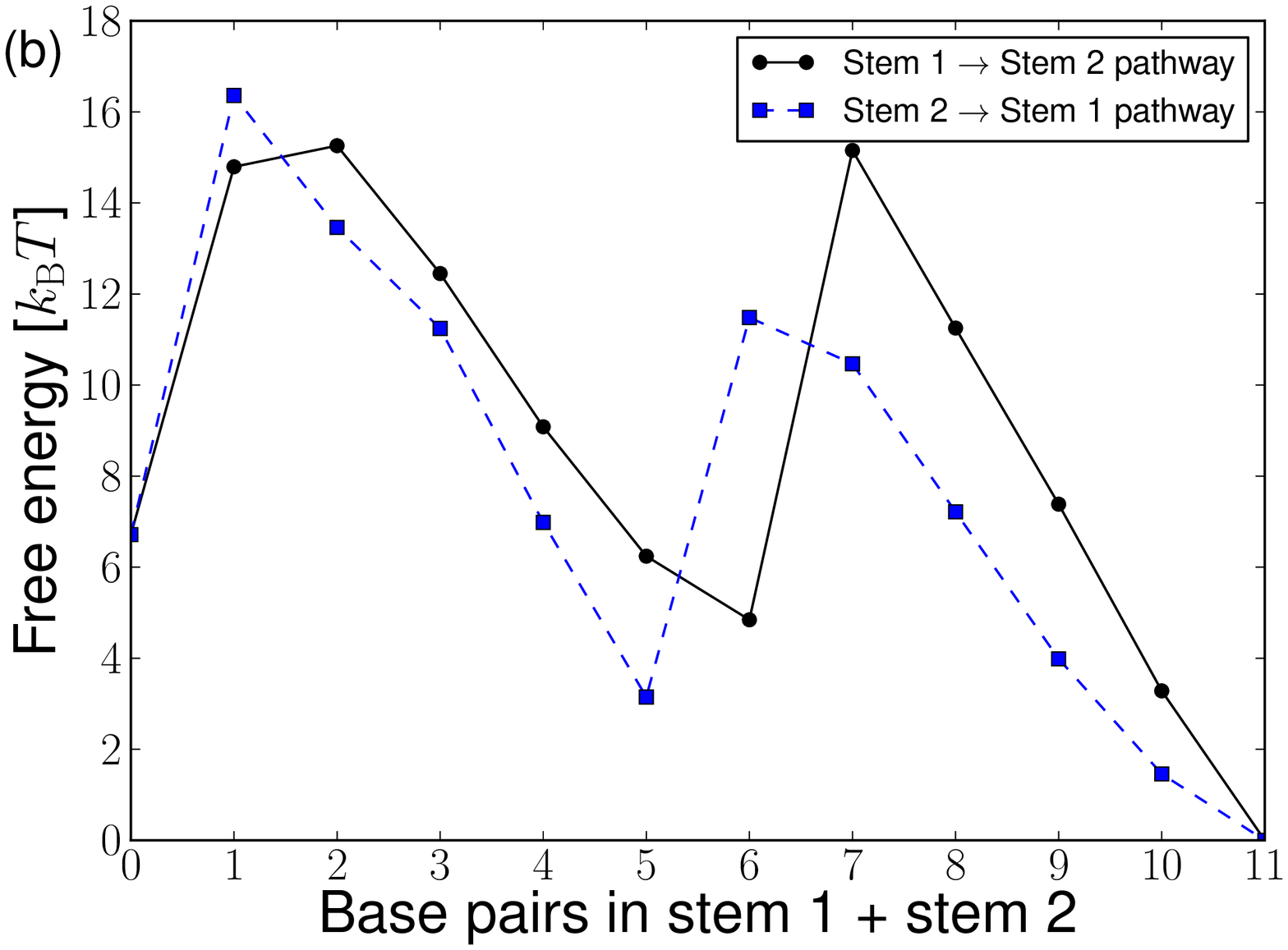}
\centering
\caption{(a) A free-energy landscape for pseudoknot formation at $48\,^{\circ}{\rm C}$. White lines denote minimum free-energy pathways.
(b) Free-energy profiles along the paths indicated in the free-energy landscape of (a). Dashed and solid lines correspond in both pictures. Only native base pairs contribute to the order parameters.
}
\label{fig_pseudo_heat}
\end{figure}

To study the thermodynamics of the
system, we ran VMMC simulations of oxRNA for $3.4 \times 10^{11}$ steps at $75\celsius$. Umbrella sampling, using the number of base pairs in each of
the pseudoknot stems as order parameters, was employed to enhance thermodynamic
sampling. We also used histogram reweighting to extrapolate our results to other temperatures. The occupation probabilities of 
the unfolded state, a single hairpin with stem 1 or stem 2 (denoted as hairpin 1 and hairpin 2),
 and the pseudoknot are shown in Fig.~\ref{fig_pseudo_yield_cv}(a).
Our simulations also allow us to extract the heat
capacity $C_{\rm V}$ from the expression 
\begin{equation}
 C_{\rm V} = \frac{\partial \left\langle U \right\rangle}{\partial T}
\end{equation}
where we use a cubic interpolation to our simulation data for $\langle U
\rangle$ in order to compute the derivative with respect to $T$. The results are
shown in Fig.~\ref{fig_pseudo_yield_cv}(b).

The experimentally measured $C_{\rm V}$ at $1\,{\rm M}$ 
\sodium  \,  has two peaks, one at $73.5\celsius$ and the other at
$95.0\celsius$.\cite{theimer2000contribution}
It was hypothesized that the two peaks correspond to the transition from an
unstructured strand to a hairpin structure and a second transition from a
hairpin structure to the full pseudoknot. Analysis of our yields supports this
claim, showing a pseudoknot to hairpin 2 transition at 
$67.7\celsius$ and transition from hairpin 2 to a single strand with no bonds in stem 1
or stem 2 at $84.8\celsius$. 
The higher temperature transition coincides with a peak in the heat capacity, whilst the lower temperature transition gives 
rise to a shoulder. While our simulations reproduce qualitatively the behavior of
the experimental system, the position of the transitions is not exactly the same as the 
 ones measured experimentally. This is not surprising, as we have shown in
Section~\ref{sec_thermo} that the model generally underestimates the melting temperatures of
hairpins.

It is of further interest to analyze the free-energy landscape of the system (Fig.~\ref{fig_pseudo_heat}).
Perhaps unsurprisingly, our results suggest that the
minimum free-energy pathway for folding this pseudoknot from a single strand
involves first forming one of the stems (forming stem 2 first is more likely as it is more stable and
has a higher yield at the considered temperature) and then closing the second
stem, instead of simultaneously forming both of them. We have previously seen similar pathways for a DNA pseudoknot.\cite{Doye13} 

One subtlety concerns the formation of the GU base pair between the seventh and thirty-fourth nucleotide. The NMR study\cite{shen1995structure} did not observe the presence of this GU base pair  in the pseudoknot structure. However, in our simulations, we find some structures where this base pair forms (thus extending the size of stem 1 from six to seven base pairs), although it has a $5\, k_{\rm B} T$ free energy penalty at $48\,^{\circ}{\rm C}$ with respect to a pseudoknot state which had only six bases in stem 1. Including this additional base-pair within the definition of stem 1 had only a small effect on the calculated yields (the positions of the equal yields points changed by less than $0.3\celsius$) and we saw at most $0.5\,k_{\rm B} T$ free-energy change for the folding pathways. We thus did not include this extra base pair in the definition of stem 1. 

The experimental NMR study\cite{shen1995structure} of the structure of the MMTV pseudoknot found that the two stems of the pseudoknot are bent with respect to each other at about $112^{\circ}$, and the AA mismatch between the sixth and the fourteenth nucleotides to be not stacked.  As can be seen in Fig.~\ref{fig_pseudo}, in oxRNA, this mismatch is typically stacked leading to an angle closer to $160^{\circ}$.  We think that this stacking of the stems reflects the overestimation of the stability of mismatches in simpler motifs (see Table \ref{table_ssmotifs}).

In summary, our model is able to describe the thermodynamics of the pseudoknot folding and predict the stabilities of the two stems, supporting the hypothesis that the peak in heat capacity at higher temperature corresponds to the pseudoknot to hairpin 2 transition. The overall secondary structure of the pseudoknot is correct in our model, which also helps to understand the tertiary structure even though we found the angle between the two stems to be larger than the one reported from experiment.

\subsection{Kissing hairpin complex}

\begin{figure}
\centering
\includegraphics[width=0.5\textwidth]{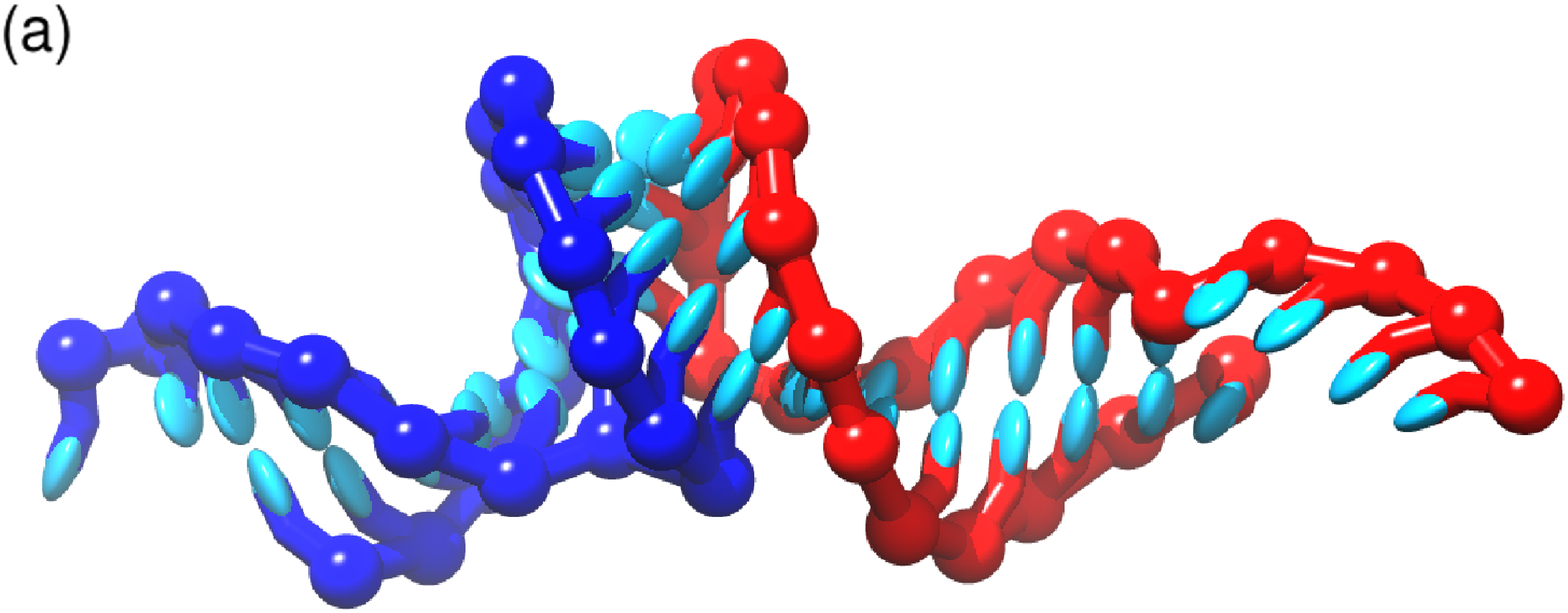}\\
\includegraphics[width=0.5\textwidth]{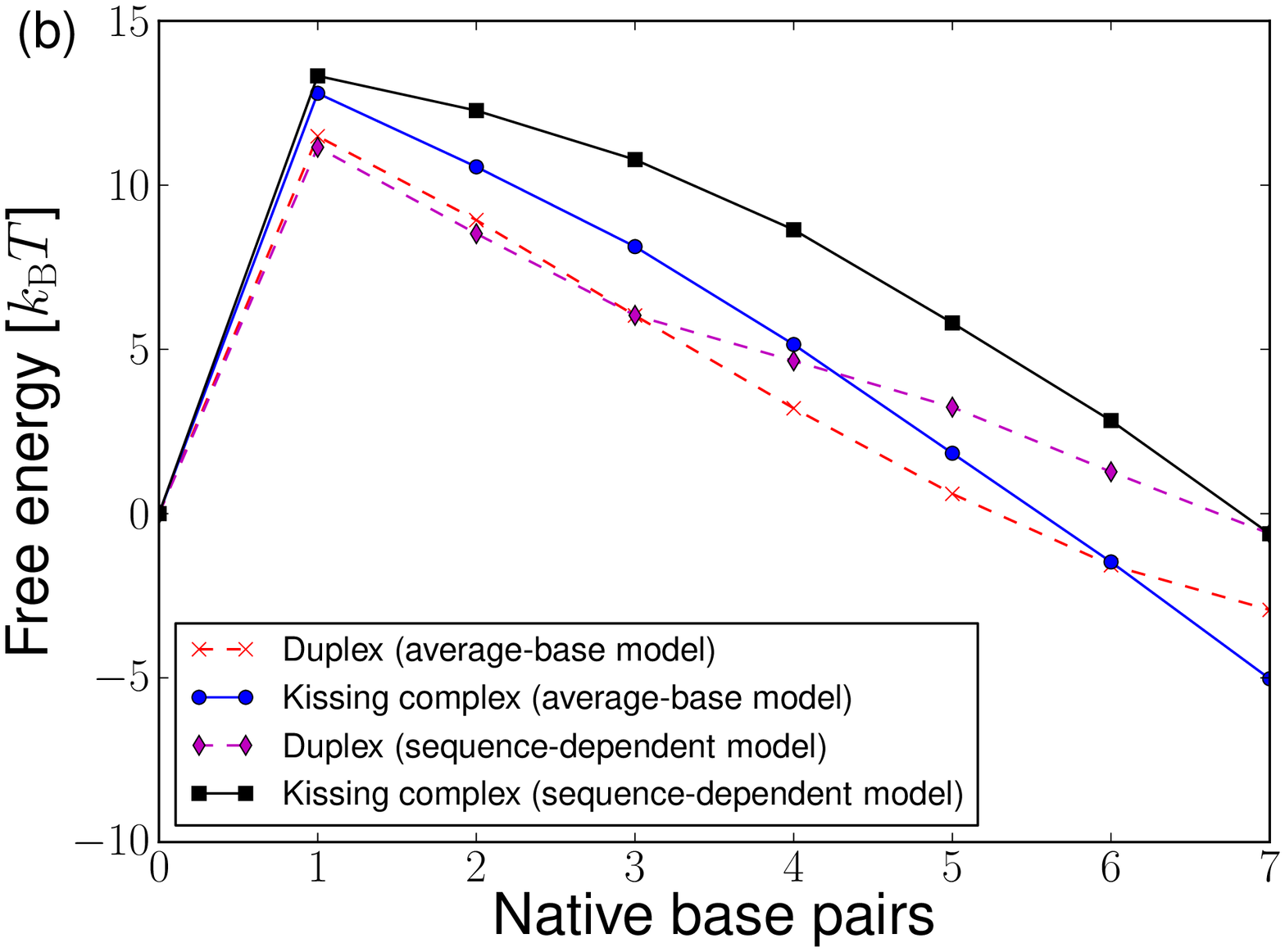}
\caption{ (a) A typical configuration of a kissing complex between two hairpins that have a complementary 7-base loops.
(b) Free-energy profiles at $45\celsius$ for forming the kissing complex and a 7-bp duplex with the same sequence as the hairpin loops. Results are shown for 
the average-base and the sequence-dependent parametrization of oxRNA.}
\label{fig_kissprofile}
\end{figure}

\begin{figure}
\includegraphics[width=0.35\textwidth]{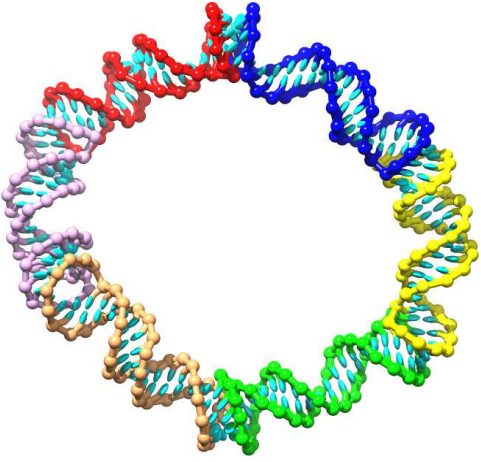}
\centering
\caption{The hexagonal RNA nanoring of Ref.~\onlinecite{Shapiro2007}, as represented by oxRNA. The structure is composed of six strands, with a total of 264 nucleotides, connected by kissing loops.}
\label{fig_nanoring}
\end{figure}

A kissing complex is a naturally occurring motif in RNA
structures\cite{Elliott11} and consists of two hairpins that have complementary
loops and can thus bind to each other. An example of such a complex as represented by oxRNA is
shown in Fig.~\ref{fig_kissprofile}. The kinetics and thermodynamics of forming an
RNA kissing complex with 7 bases in the loops was experimentally studied in
Ref.~\onlinecite{salim2012thermodynamic} at varying salt concentrations,
including 1\,M \sodium, the concentration at which our model was parametrized.

To examine the capability of oxRNA to describe kissing complexes, we studied the
melting of this kissing complex using both the average-base and
the sequence-dependent parametrization of oxRNA and found the transition at a point which is approximately consistent with the observed experimental behavior. The 7-base loops in the two
hairpins have fully complementary sequences ($5^{\prime}$-GGAAAUG-$3^{\prime}$ and its Watson-Crick complementary sequence). All melting
simulations were run in a volume corresponding to an equal strand concentration of $3.36 \times 10^{-4}\,\rm{
M}$. 

For the average-base representation we found the melting temperature of the kissing hairpins to be $62.7\celsius$ which compared to $53.6\celsius$ for a 7-bp duplex with the same sequences as the loops. For the sequence-dependent
model, we found the melting temperature of the kissing complex to be
$44.8\celsius$, similar to $45.2\celsius$ for this 7-bp duplex. 
The free-energy profiles for both average-base and sequence-dependent models at $45\celsius$ are shown in Fig.~\ref{fig_kissprofile}. 

For most sequences, we find that the kissing hairpin loop is more stable than the separate duplexes with respect to the unbound state. 
We find that with increasing temperature, the kissing complex loses less stability with respect
to the unbound state than a duplex at the same temperature and strand
concentration. This trend 
can be rationalized in terms of the fact that a constrained loop loses less 
configurational entropy upon
binding than an unconstrained single strand does.  Furthermore, the kissing
complex also gets an extra enthalpic stabilization from a coaxial stacking
interaction between the loop and the stem nucleotides.  These two effects help explain why, on average, the kissing hairpins are more stable, especially at higher temperatures.  However, the kissing hairpins do not satisfy the enthalpic contributions as well as a perfectly formed duplex does.  Thus, at low temperature, the duplex can be more stable.  Which of these effects dominates depends on the sequence, and if the melting happens before the switch of which motif is more stable, then the duplex will have a higher melting temperature, which we find for a minority of sequences at our strand concentration. For the sequence above, the melting temperatures are very close. 
Note that the hairpin loops are sufficiently short 
that kissing complex formation is not inhibited by the topological requirement that the linking number between the loops must remain zero. This contrasts with
previous simulations of DNA kissing complexes between hairpins that have 20-base complementary loops.\cite{Romano12a}

The thermodynamics of this kissing complexes was studied in Ref.~\onlinecite{salim2012thermodynamic} using isothermal titration calorimetry (ITC) at 1\,M \sodium.
Up to $35\celsius$ the authors found evidence of a transition to a kissing complex after the injection of the reactants, but did not observe a transition at $45\celsius$.  The authors claim to measure only a $0.6$ kcal/mol change in the $\Delta G$ for forming the kissing complex between $10\celsius$ and $35\celsius$ while observing a significant increase ($19.5$ kcal/mol) in $\Delta H$ along with a compensating increase in $\Delta S$. Such behavior is not observed in our simulations, where we see a classic quasi-two-state transition in which $\Delta H$ and $\Delta S$ change slowly with temperature, similarly to a duplex association. If the yields of the kissing complexes in our simulations were extrapolated to the concentrations used in the experiment, we would predict a yield of 36\% at $35\celsius$ and 5\% at $45\celsius$. 
We note that the thermodynamic parameters in Ref.~\onlinecite{salim2012thermodynamic} were derived with the assumption that the transition was fully saturated after the injection of the reactants in the ITC experiment, which is incompatible with the experimentally inferred value of $\Delta G$. If the transitions were in fact not fully saturated, then it is possible that the experiments are consistent with a more typical quasi-two-state transition as observed in our model with a melting temperature approximately consistent with that found by us.

It is also interesting to use oxRNA to probe the structure of this kissing complex because it is a motif
that has been used in RNA nanotechnology.
Molecular dynamics simulations of the kissing complex 
using an all-atom representation (Amber) measured the
angle between the helical stems at $300\,{\rm K}$ and $0.5\,$M monovalent salt to be
approximately 120$^{\circ}$.\cite{Shapiro2007} 
Based on this finding, a hexagonal ring
nanostructure that can be assembled from six RNA building blocks was proposed. Each of the
proposed building blocks is an RNA strand that in the folded state has a stem and
two hairpin loops. The sequences in the loops are designed to bind to the
complementary block to allow the assembly of a hexagonal ``nanoring'' via the kissing complex interaction. This computationally proposed RNA nanoring design was
later experimentally realized by self-assembly.\cite{grabow2011self} The
nanoring can be functionalized by including siRNA sequences either in the
hairpin stems or by appending siRNA sequences onto the stems. Experiments in
blood serum have shown that the nanoring protects the loop
regions of the assembly blocks from single-strand RNA endonucleolytic cleavage,
making the nanoring a promising tool for \textit{in vivo} nanotechnology
applications.\cite{grabow2011self} 

Simulations of oxRNA at $25\celsius$ allowed us to measure the
angle between the helical axes of the hairpin stems. 
We found this angle to fluctuate around an average value of $133.9^{\circ}$ with
a standard deviation of $14.8^{\circ}$. Hence, an octagon would probably be the most relaxed nanoring for oxRNA, and therefore favored by enthalpy. Smaller 
rings would be favored by translational entropy.

To illustrate the capabilities of our oxRNA model, we constructed the hexagonal RNA nanoring 
of Ref.~\onlinecite{Shapiro2007} (Fig.~\ref{fig_nanoring}) by starting a simulation with
six folded hairpins blocks and introducing a harmonic potential between
the complementary loop regions, which helped the kissing interactions to form. Once the 
nanoring was completed, the harmonic traps were removed and the assembled structure
was relaxed in a molecular dynamics simulation.
We found the angle between the stems of the kissing hairpins in the nanoring to fluctuate around a mean of 
$124.9^{\circ}$. The thermal fluctuations around the mean value had a standard deviation of $14.4^{\circ}$, which is similar to that of the single kissing complex.
It would also be possible to use our model
to study the mechanical properties of the nanoring, as well as
the thermodynamics and kinetics of its self-assembly from the building blocks, but
such a study is beyond the scope of this article.

A typical relaxation time for the angle between adjacent kissing complexes or the energy self-correlation function of the assembled nanoring structure corresponds to about a minute of CPU time or less on a standard 2.2 GHz processor. This example shows that structural investigations of systems of the order of hundreds of bases are well within the capabilities of the oxRNA model using a single CPU, and if multiple CPUs are used, or a GPU chip is used,\cite{rovigatti2014comparison} then structural properties and fluctuations around equilibrium can be studied for systems on the order of thousands of base pairs, as can be done for oxDNA.\cite{Doye13}

\subsection{Hairpin unzipping}

\begin{figure}
\includegraphics[width=0.49\textwidth]{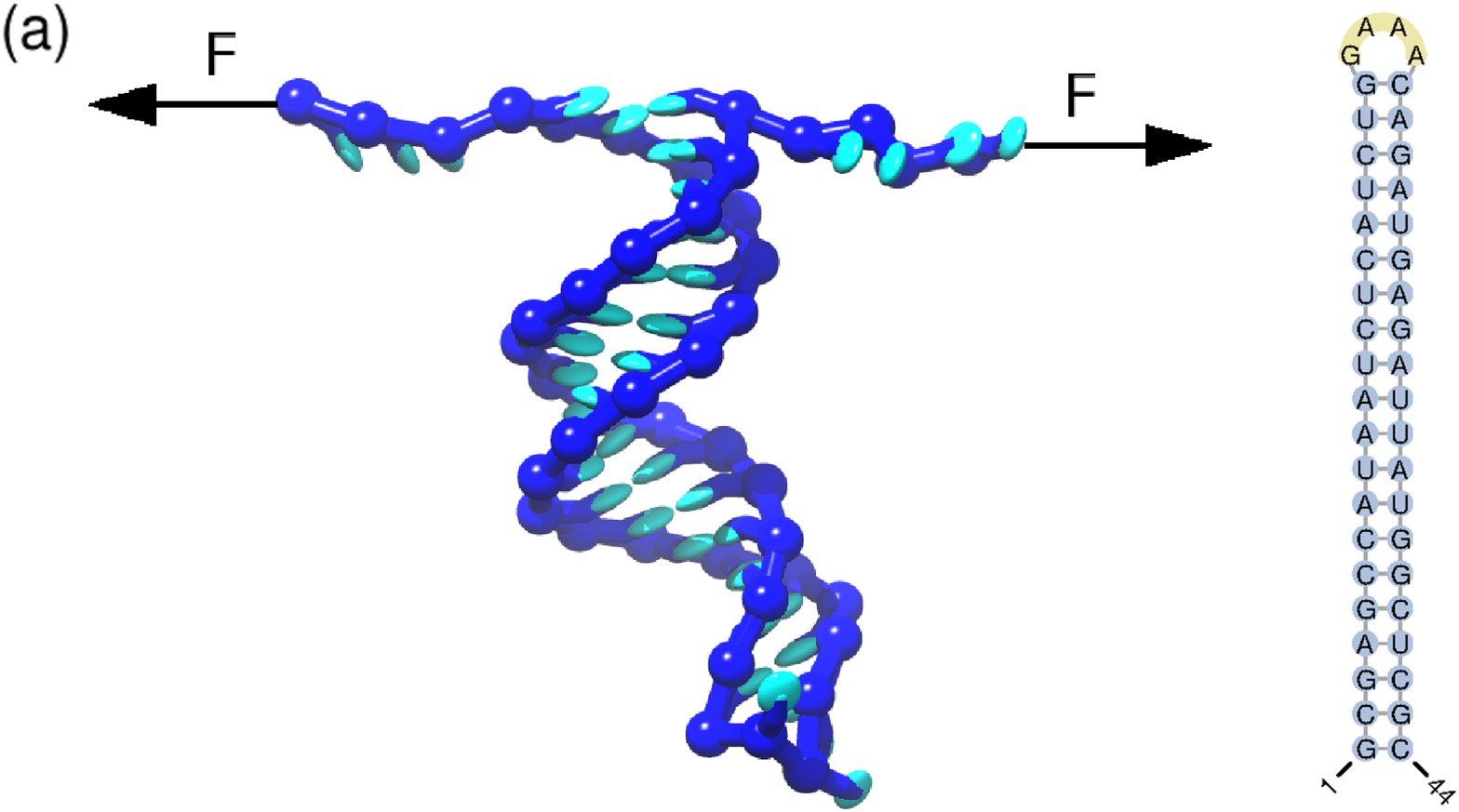}\\
\includegraphics[width=0.5\textwidth]{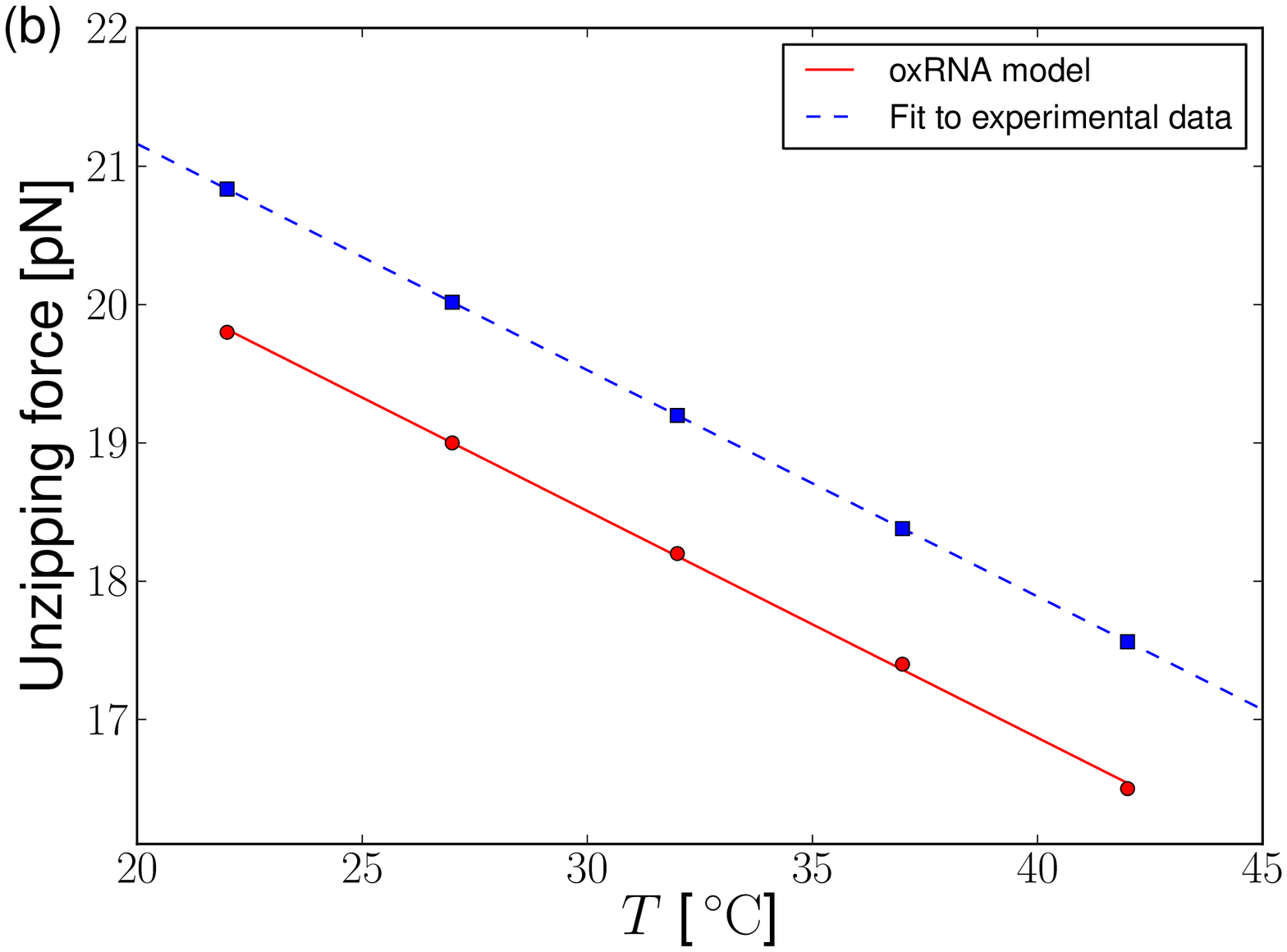}
\centering
\caption{(a) A snapshot from an oxRNA simulation of unzipping of the CD4 hairpin and a schematic representation of the CD4 hairpin sequence created with the PseudoViewer software.\cite{byun2006pseudoviewer}
 (b) The unzipping forces for the CD4 hairpin as a function of temperature for the oxRNA model (full circles) along with a linear fit to the data. For comparison, the fit to the 
experimentally measured unzipping force at 1\,M \sodium \  at temperatures $22$, $27$, $32$, $37$ and $42\celsius$ is also shown.}
\label{fig_pull}
\end{figure}

RNA hairpin unzipping has been used in
experiment to study the thermodynamics of base pairing and the mechanical properties of RNA,
with the kinetics of the process also being of interest.
\cite{Manosas2006,bizarro2012non,stephenson2013combining} Unzipping the same
sequence under different salt and temperature conditions 
can provide systematic data on the physical properties of RNA that, for example, could be used
to improve
the parametrization of thermodynamic models of RNA.

Here we consider the unzipping of the CD4 hairpin (shown schematically in Fig.~\ref{fig_pull}(a)), which has a 20-bp stem and 4 bases in the loop. It was studied experimentally by pulling at different rates and measuring the unzipping force\cite{Manosas2006,bizarro2012non,stephenson2013combining} for each trajectory.
While it is possible to simulate pulling the hairpins ends at a given rate in the oxRNA model, direct comparisons with experimentally observed unzipping forces are difficult because for a coarse-grained model there is not a straightforward way to map the simulation time to the experimental one. Furthermore, to obtain an estimate of the average unzipping force for a given pulling rate, one needs to average over multiple trajectories, and generating such trajectories can be quite time consuming, especially for very slow pulling rates.

A more suitable experimental setup for comparison with a coarse-grained simulation comes from Ref.~\onlinecite{stephenson2013combining}, where the authors performed force-clamp experiments at different temperatures and salt concentrations. In the experimental setup, they kept the forces applied to the first and last base of the hairpin constant and measured the folding and unfolding rate of the hairpin, from which they inferred the free-energy difference between the open and closed state. By interpolating results for a range of applied forces, they were able to obtain the unzipping force, which was defined as the force for which the free-energy difference was zero. 

To compare to the experimental results, we ran VMMC simulations of the CD4 hairpin with a constant force of $19.7\,{\rm pN}$ applied to the first and last nucleotide at $23\celsius$. We then extrapolated the free energies of the open and closed state by the histogram reweighting method to forces ranging from $16.2$ to $20.7\,{\rm pN}$ and to the temperatures at which the experiments were carried out ($22$, $27$, $32$, $37$ and $42 \celsius$). We only allowed bonds between the native base pairs to avoid sampling of metastable secondary structures that would slow down our simulations. We considered the hairpin to be closed if at least one of the bonds in the stem was present. For each temperature considered, we performed a linear interpolation of the free-energy difference between closed and open state as a function of force to obtain the unzipping force for which the difference is 0. The unzipping forces we obtain are shown in Fig.~\ref{fig_pull}, along with a fit. We also show for comparison the fit to the experimentally observed unzipping forces\cite{stephenson2013combining} at $1\,$M \sodium, expressed in the form  
\begin{equation} 
\label{linfit}
 F_{\rm unzip}(T) = a - c T,
\end{equation}
 where $F_{\rm unzip}$ is the unzipping force at temperature $T$. The values obtained in the experiment\cite{stephenson2013combining} were $a = 69.1\,{\rm pN}$ and $c = -0.164\,{\rm pN / K}$. Fitting Eq.~\ref{linfit} to our simulation data, we obtained the same value for $c$ and $68.2\,{\rm pN}$ for $a$. The values of the fitting parameters varied by at most $6\%$ between the fits to unzipping forces obtained from three independent sets of generated states. Thus, oxRNA is able to reproduce the unzipping force with 5\% ($1\,{\rm pN}$) accuracy and fully captures the trend with temperature.
  
  It would be of further interest to study the force-extension properties of a hairpin which contains various secondary structure motifs such as bulges and internal loops or which has regions with variable sequence strengths (such as AU rich and GC rich regions). However, such a study is beyond the scope of this article.

\section{Summary and Conclusions}
We have presented a new off-lattice coarse-grained model for RNA,  oxRNA, which aims to capture basic thermodynamic, structural and mechanical properties of RNA structures with a minimal representation 
and pair-wise interaction potentials.  OxRNA is  specifically developed to allow for efficient simulations of large structures (up to hundreds of nucleotides on a single CPU), and for reactions involving multiple strands of RNA, which are important for applications in RNA nanotechnology. 
   Our coarse-graining strategy is closely linked to our previous successful coarse-graining of DNA with oxDNA.\cite{Ouldridge2011}  Rather than focusing mainly on reproducing structure, as many other previous models have done, here we tried to systematically compare to a whole suite of different properties. 
   
We employed a ``top-down'' coarse-graining approach, where we aim to reproduce free-energy differences between different states (such as opened and closed state of a hairpin) as measured in experiment. 
OxRNA represents each
nucleotide (i.e., sugar, phosphate and base) as a single rigid body with multiple
interaction sites.  The model is capable of representing RNA structures such as
hairpins and duplexes and is designed to reproduce the A-helical form of duplex
RNA.  We used a histogram reweighting method~\cite{oxDNA} to parametrize the
model to reproduce the thermodynamics of short duplexes and hairpins.
Currently, the oxRNA model captures Watson-Crick and wobble base-pairing interactions as well as various types of stacking interaction.  However it was not designed to capture non-canonical interactions such as Hoogsteen or sugar-edge hydrogen-bonded base pairs, or ribose zippers.  Nevertheless, it can reproduce some important tertiary interaction motifs, in particular coaxial stacking of helices, kissing loop interactions, and pseudoknots.

The model is currently parametrized for a salt concentration of $1\,{\rm M}$, as this corresponds to the 
conditions for the melting experiments used for the nearest-neighbor model to which our model was
parametrized. Explicit
electrostatic interactions are not included,  because they are very short-ranged at
high salt and thus can be incorporated into the short-ranged excluded volume terms in the potential.
This  excluded volume also prevents a strand from crossing itself or other strands, forbidding topologically impossible trajectories in kinetic simulations. 
 It is possible to use the same coarse-graining techniques to parametrize the model at lower salt concentrations.  However, as the screening lengths become longer, different longer-ranged forms for the interactions may need to be used to capture the correct physics. 
We note however that nanotechnology experiments {\it in vitro} are usually carried out 
in high salt conditions.

To test our model, we investigated the
thermodynamics of short duplexes and hairpins with different sequence content,
as well as various other secondary structures such as bulges, internal and terminal
mismatches. 
We found that in comparison with our previous model for DNA, parametrizing RNA thermodynamics is a more challenging task. For example, experimental melting temperatures of a duplex of a given length can differ by as much as $70\celsius$ between weak and strong sequences with Watson-Crick base pairs, as opposed to $50\celsius$ for DNA.\cite{oxDNA} So sequence effects are stronger in RNA. Including wobble base pairs presents further challenges, as some base pair steps that include two or more neighboring wobble base pairs have a positive contribution to the free energy of duplex. Although it is not possible to capture this effect with the current representation of our model, adding the structural effects of wobble base pairs on the double helix may provide means to improve this aspect of the model. 
Finally, we found that oxRNA overestimates the stability of duplexes with mismatches in internal loops. This could lead to an overestimation of the stability of misfolded structures and complicate the study of the folding of RNA strands that have multiple metastable states with mismatches. Nevertheless, even though oxRNA does not reproduce the exact melting temperatures for structures with internal mismatches and bulges, we do observe, as expected, a decrease in melting temperatures of a duplex with internal mismatches or bulges with respect to the fully complementary strands. The observed changes in melting temperatures are within the orders of magnitude as predicted by the NN-model for the same motifs and capture correctly the direction of the change.

We have tested the mechanical properties of the RNA duplex as represented by the model and found its persistence
length to be about half of the reported persistence length of RNA molecules at
high salt conditions. The model hence has the correct order of magnitude for the duplex persistence length and provides sufficiently accurate mechanical behavior for most applications, where
individual double helical sections are likely to be much shorter than the persistence length.

In Section \ref{sec_examples}, we provided some  applications of the model that illustrate its strength and potential utility. 
In particular, we investigated the thermodynamics of pseudoknot folding and
the thermodynamics and structure of a kissing hairpin complex. 
We also
showed that oxRNA can be used to model large nanostructures like an RNA nanoring
composed of 264 nucleotides on a single CPU. The computational cost of oxRNA is very similar to oxDNA where simulations of a DNA origami motif with 12\,391 nucleotides have been achieved.\cite{Doye13} Finally, our model is able to reproduce experimental results for the mechanical unzipping of a hairpin quite
closely, to within an accuracy of $1\,\rm{pN}$, illustrating oxRNA's potential to study mechanical properties of
RNA constructs. 

Although we did not describe applications with detailed dynamics (the simulations are typically 
quite involved and so beyond the scope of this study), we want to emphasize that oxRNA is particularly well-suited for such tasks. For example, oxDNA has been used to study the detailed dynamics of hybridization,\cite{ouldridge2013dna} toehold-mediated strand displacement\cite{srinivas2013biophysics} and hairpin formation.\cite{Doye13} Studying similar processes would be very feasible for oxRNA. For example, it should be possible to use the model to obtain estimates 
of the rates of a strand displacement reactions as a function of length of the toehold as well as temperature. OxRNA can further be used to study the self-assembly of nanostructures such as RNA nanorings and to investigate both the thermodynamics and the kinetics of such systems. Although the model is currently only parametrized at high salt concentration, oxRNA can be also used to qualitatively study processes of biological relevance, for instance, the folding pathways of RNA strands or {\em in vivo} applications of nanotechnology.

 At this point it is interesting to reflect on some similarities and differences between the coarse-graining of oxDNA and oxRNA.  Although oxRNA can clearly reproduce a good number of properties of RNA, quantitative differences with experiment for the melting temperatures of certain motifs are larger than they are for oxDNA.  Moreover, it was more difficult to simultaneously reproduce the thermodynamics and the correct persistence length or the force-extension curves. One reason for these differences may be that RNA itself exhibits more complex behavior than DNA, and so is harder to coarse-grain.  It is intuitively obvious that the compromises made to increase tractability mean that not all properties of the underlying system can be simultaneously captured by a more simplified description, a general phenomenon  that has been called ``representability problems".\cite{Louis2002,Louis2010no} One consequence of representability problems is that in general, fitting too strongly to one set of input data (say structure, as is often done for other RNA models) will often lead to larger errors in other quantities, say thermodynamics.  We tried to compromise between different requirements for oxRNA.  However,  in order to make further progress, more detailed fitting may not be enough. Instead a more complex and most likely less tractable representation of the full interactions may need to be chosen.   For example, for RNA it remains to be seen whether our single  nucleotide-level model can be easily extended to generate a better representation of both structure and thermodynamics, or whether, say, a more complex model is needed to achieve the next level of accuracy.    Clearly, including  electrostatic effects for lower salt-concentrations, or implementing tertiary structure contacts, for example non-canonical base pairing interaction (such as Hoogsteen base pairs) and hydrogen bonding between  a sugar group and a base will also need an extension of the current representation.  

 Given the challenges and complexity of RNA modelling, it is unsurprising that oxRNA performs less well than oxDNA for the whole ensemble of motifs we considered. 
 However, we believe that it is a non-trivial achievement to create a model that can semi-quantitatively reproduce such a wide range of the thermodynamic data. 
 The properties of our model have been comprehensively tested on a variety of systems, ranging from secondary structure motifs to systems such as kissing complexes and hairpin unzipping. 
 Our model is also currently the only RNA model with reported mechanical properties, which were tested by measuring its persistence length, force-extension curve and duplex overstretching.

Finally, the simulation code implementing molecular dynamics and Monte Carlo algorithms
for our oxRNA model is available for download at dna.physics.ox.ac.uk.

\section*{Acknowledgments} 
The authors thank Agnes Noy for helpful discussions and Lorenzo Rovigatti for his help with the simulation code development.
The authors acknowledge the Engineering and Physical
Sciences Research Council and University College (Oxford) for financial support and the Advanced Research Computing, University of Oxford for
computer time. P.~\v{S}.~is grateful for the award of a Scatcherd
European Scholarship.

\renewcommand{\appendixname}{Supplementary Material}
\appendix

\section{}
 \setcounter{figure}{0}
 \makeatletter 
 \renewcommand{\thefigure}{A\@arabic\c@figure}
 \setcounter{equation}{0}
 \renewcommand{\theequation}{A\@arabic\c@equation}
 \setcounter{table}{0}
 \renewcommand{\thetable}{A-\@Roman\c@table}

\subsection{Fitting of the helical axis of a duplex}
\label{app_axis_fit}

\begin{figure}
\includegraphics[height=0.25\textheight]{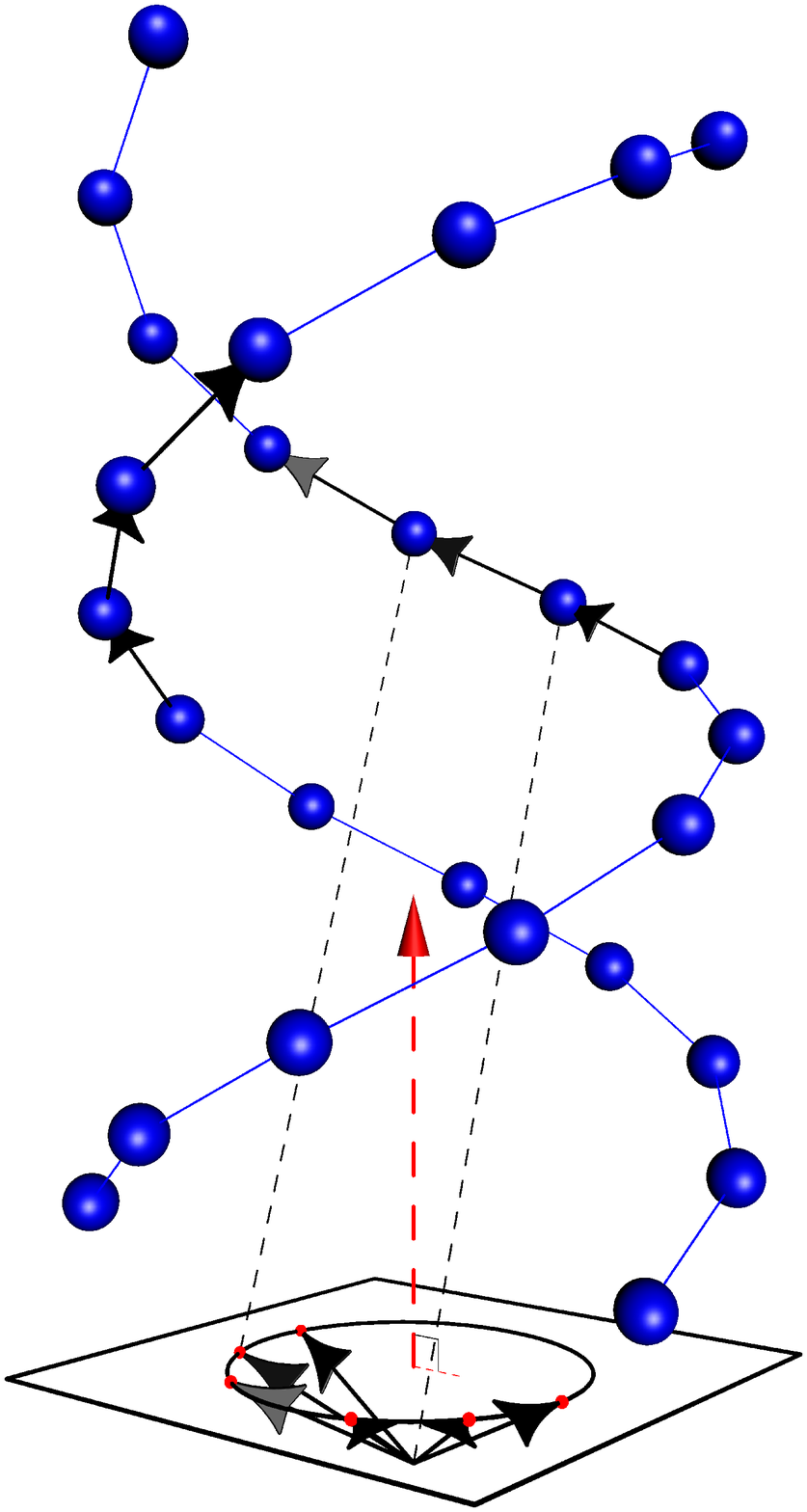}
\centering
\caption{Fitting a helical axis to the duplex. The blue spheres show the positions of the backbone sites. The arrows represent vectors pointing from a nucleotide backbone site to its neighbor's backbone site. When the vectors are placed onto the same origin, their endpoints would lie in a plane for the case of a perfect A-helical structure. A plane can hence be fitted through the endpoints of these vectors. A vector perpendicular to this plane is used as the helical axis (red dashed arrow).}
\label{fig_axis_example}
\end{figure}

We discussed oxRNA's description of the structure of the A-helix in Section \ref{sec_structure}. For the considered 13-bp long duplex, we fitted the helical axis in the following way.
For each base in the first strand, we took the vector pointing from its backbone site to the backbone site of its $3^{\prime}$-neighbor and for each base in the second strand, 
we considered the vectors pointing to the $5^{\prime}$-neighbor's backbone site. For a perfect A-helical structure, the endpoints of all the vectors would all lie in the same plane
if the origins of the vectors were all placed at the same point. The structure of the duplex is subject to thermal fluctuations and hence the 
plane has to be fitted through the endpoints of the vectors. The first and last two base pairs
were not included in order to exclude end effects. The vector perpendicular to the plane was then taken to be the helical axis. The fitting of the helix is schematically illustrated in Fig.~\ref{fig_axis_example}.

\subsection{The potentials and nucleotide representation in the RNA model}
\label{appendix_potentials}
The model and its potentials were introduced in Section \ref{sec_model_and_par} and here we provide a detailed description of the interaction potentials and the nucleotides.
We first describe the representation of each nucleotide as a rigid body in Section \ref{app_rep} and then give the explicit expression for each of the potential terms in Section \ref{Functional_forms}. All the values of the potential parameters are in an internal unit system of the downloadable simulation code where 1 distance unit = $8.4\,\angstrom$ and 1 energy unit = $41.4\, {\rm pN\,nm} = 10\, k_{\rm B} T$ for $T=300\,$K. In molecular dynamics simulations, we set $1$ mass unit to correspond to the average weight of a nucleotide, $321.4$ AMU, which gives us the simulation time unit corresponding to $3.06 \times 10^{-12}\,{\rm s}$ in SI units.

\subsubsection{Representation}
\label{app_rep}
Each nucleotide in the oxRNA model is represented as a single rigid body with multiple interaction sites. Each nucleotide has backbone, hydrogen-bonding, cross-stacking and $3^{\prime}$- and $5^{\prime}$-stacking interaction sites. The position and orientation of each nucleotide is uniquely specified by its center of mass position and the perpendicular unit vectors $\mathbf{a}_3$ and $\mathbf{a}_1$, where $\mathbf{a}_1$ is a unit vector pointing from the center of mass to the hydrogen-bonding site and $\mathbf{a}_3$ is defined in Fig.~\ref{fig_nuc}. In a duplex configuration, the $\mathbf{a}_3$ vector would be pointing towards the $5^{\prime}$-neighboring nucleotide. We further define $\mathbf{a}_2 = \mathbf{a}_3  \times \mathbf{a}_1$. The nucleotide as represented by oxRNA is schematically shown in Fig.~\ref{fig_nuc}. The small colored circles indicate the position of the interaction sites, while the large circles around hydrogen bonding and backbone sites indicate the interaction radius of the excluded-volume interactions. 

The interaction potentials are functions of the distances between the relevant interaction sites as well as the angles between intersite vectors and the respective orientation vectors $\mathbf{a}_3$, $\mathbf{a}_1$, $\mathbf{p}_{3^{\prime}}$ and $\mathbf{p}_{5^{\prime}}$, where we define
\begin{eqnarray}
 \mathbf{p}_{3^{\prime}} &=& -0.46 \mathbf{a}_1  -0.53 \mathbf{a}_2 + 0.71 \mathbf{a}_3 \label{A_p3} \\
 \mathbf{p}_{5^{\prime}} &=& -0.1 \mathbf{a}_1 - 0.84 \mathbf{a}_2 + 0.53 \mathbf{a}_3 . \label{A_p5}
\end{eqnarray}

We also define the following vectors which are then used in the definitions of the potentials in the oxRNA model (Eq.~\ref{eq_potential}):
  \begin{itemize}
 \item $\delta \mathbf{r}_{\rm backbone}$: the vector between backbone sites of the nucleotides. If the nucleotides are nearest neighbors, it is pointing towards the nucleotide's $3^{\prime}$-neighbor's backbone site
 \item $\delta \mathbf{r}_{\rm HB}$: the vector between the hydrogen-bonding sites of the interacting nucleotides, pointing from the first nucleotide towards the second one
 \item $\delta \mathbf{r}_{\rm coaxial~st.}$: the vector between the coaxial stacking sites of the interacting nucleotides, pointing from the first nucleotide towards the second one
 \item $\delta \mathbf{r}_{\rm stack}$: the vector pointing from the $3^{\prime}$-stacking site of a nucleotide to the $5^{\prime}$-stacking site of its $3^{\prime}$-neighbor
 \item $\delta \mathbf{r}_{\rm back-base}$/$\delta \mathbf{r}_{\rm base-back}$: the vector pointing from the backbone/hydrogen-bonding site of the first nucleotide to the hydrogen-bonding/backbone site of the second nucleotide  
\end{itemize}

We further define the following angles that are used in the potential functions: 
\begin{eqnarray} 
  \theta_{1} &=& \arccos \left( \mathbf{a}_1  \cdot \mathbf{b}_1 \right)\\
  \theta_{2} &=& \arccos \left( -\mathbf{b}_1  \cdot \widehat{\delta \mathbf{r}}_{\rm HB} \right)\\
  \theta_{3} &=& \arccos \left( \mathbf{a}_1  \cdot   \widehat{\delta \mathbf{r}}_{\rm HB}   \right)\\
  \theta_{4} &=& \arccos \left( \mathbf{a}_3  \cdot \mathbf{b}_3 \right)\\
  \theta_{5} &=& \arccos \left( \mathbf{a}_3  \cdot    \widehat{\delta \mathbf{r}}_{\rm coaxial~st.}       \right)\\
  \theta_{6} &=& \arccos \left( -\mathbf{b}_3  \cdot   \widehat{\delta \mathbf{r}}_{\rm coaxial~st.}    \right)\\
  \theta_{5^{\prime}} &=& \arccos \left( \mathbf{a}_1  \cdot\widehat{\delta \mathbf{r}}_{\rm stack} \right)\\
  \theta_{6^{\prime}} &=& \arccos \left(-\mathbf{b}_3  \cdot\widehat{\delta \mathbf{r}}_{\rm stack} \right)\\
  \theta_{7} &=& \arccos \left( -\mathbf{b}_3  \cdot\widehat{\delta \mathbf{r}}_{\rm HB}       \right)\\
  \theta_{8} &=& \arccos \left( \mathbf{a}_3  \cdot  \widehat{\delta \mathbf{r}}_{\rm HB}      \right)\\
  \theta_{9} &=& \arccos \left( -\mathbf{p}_{3^{\prime}}  \cdot  \widehat{\delta \mathbf{r}}_{\rm backbone}      \right)\\
  \theta_{10} &=& \arccos \left( -\mathbf{q}_{5^{\prime}}  \cdot  \widehat{\delta \mathbf{r}}_{\rm backbone}     \right)\\
  \cos \left( \phi_1 \right) &=&    \widehat{\delta \mathbf{r}}_{\rm backbone} \cdot \mathbf{a}_2  \\
  \cos \left( \phi_2 \right) &=&   \widehat{\delta \mathbf{r}}_{\rm backbone} \cdot \mathbf{b}_2     \\
  \cos \left( \phi_3 \right) &=&  \widehat{\delta \mathbf{r}}_{\rm coaxial~st.} \cdot \left( \widehat{\delta \mathbf{r}}_{\rm backbone} \times \mathbf{a}_1 \right)  \\
  \cos \left (\phi_4 \right) &=&  \widehat{\delta \mathbf{r}}_{\rm coaxial~st.} \cdot \left( \widehat{\delta \mathbf{r}}_{\rm backbone} \times \mathbf{b}_1 \right) ,  
\end{eqnarray}
where we use the notation  $\mathbf{b}_1$, $\mathbf{b}_2$, and $\mathbf{b}_3$ to define the orientation vectors of the second nucleotide participating in the interaction (the orientation vectors of the first nucleotide are denoted as $\mathbf{a}_1$, $\mathbf{a}_2$, and $\mathbf{a}_3$). The vector $\mathbf{q}_{5^{\prime}}$ corresponds to the $\mathbf{p}_{5^{\prime}}$ vector of the $3^{\prime}$-neighbor of the interacting nucleotide,~i.e.\ using the same definition as in Eq.~\ref{A_p5}, but substituting $\mathbf{b}$ for $\mathbf{a}$.

\subsubsection{Potentials}
\label{Functional_forms}
The oxRNA potential consists of a sum of potential functions designed to represent different physical interactions, with some of the potentials being products of multiple
potential functions. The functions that are used in the potentials are:
\begin{itemize}
\item FENE spring (used in  $V_{\rm{backbone}}$):
\begin{equation}
V_{\rm FENE}(r,  \epsilon,r^0,\Delta) = - \frac{\epsilon}{2} \ln \left( 1- \frac{(r-r^0)^2}{\Delta^2} \right).
\end{equation}
\item Morse potential (used in $V_{\rm{stack}}$ and $V_{\rm{H.B.}}$):
\begin{equation}
V_{\rm Morse}(r, \epsilon, r^0,d) = \epsilon \big(1-\exp{(-d(r-r^0))}\big)^2.
\end{equation}
\item Harmonic potential (used in $V_{\rm{cross~st.}}$ and $V_{\rm{coaxial~st.}}$):
\begin{equation}
V_{\rm harm}(r, k, r^0) =  \frac{k}{2} \left(r - r^0 \right)^2.
\end{equation}
\item Lennard-Jones potential (used in excluded volume potentials $V^{'}_{\rm{exc}}$ and $V_{\rm{exc}}$):
 \begin{equation} 
V_{\rm LJ}(r, \epsilon, \sigma) = 4\epsilon \left[ \left({\sigma \over r} \right) ^{12} - \left( {\sigma \over r} \right) ^{6} \right].
\end{equation}
\item Quadratic modulation terms (used for angular modulation of the anisotropic potentials  $V_{\rm{H.B.}},  V_{\rm{cross~st.}},  V_{\rm{stack}}$ and $V_{\rm{coaxial~st.}}$):
\begin{equation}
V_{\rm mod} (\theta, a, \theta^0) = 1 - a (\theta-\theta^0)^2.
\end{equation}
\item Quadratic smoothing terms for truncation (used in all potentials with the exception of $V_{\rm{backbone}}$) in order to make the potentials differentiable functions that are equal to $0$ beyond some specific cutoff distance:
\begin{equation}
V_{\rm smooth} (x, b, x^c) = b(x^c - x)^2.
\end{equation}
\end{itemize}

\begin{widetext}

\begin{figure}
\includegraphics[width=0.48\textwidth]{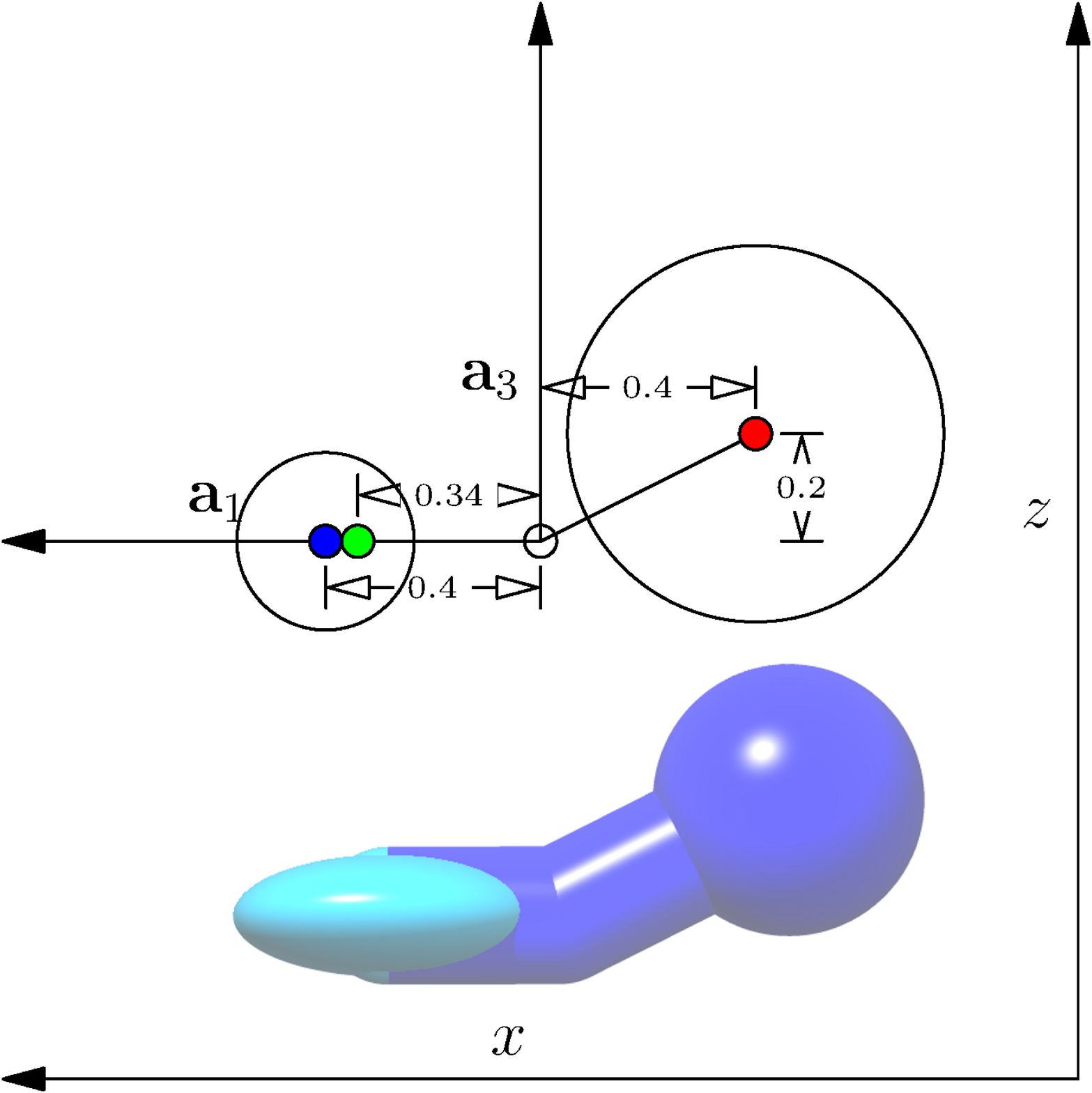}
\includegraphics[width=0.48\textwidth]{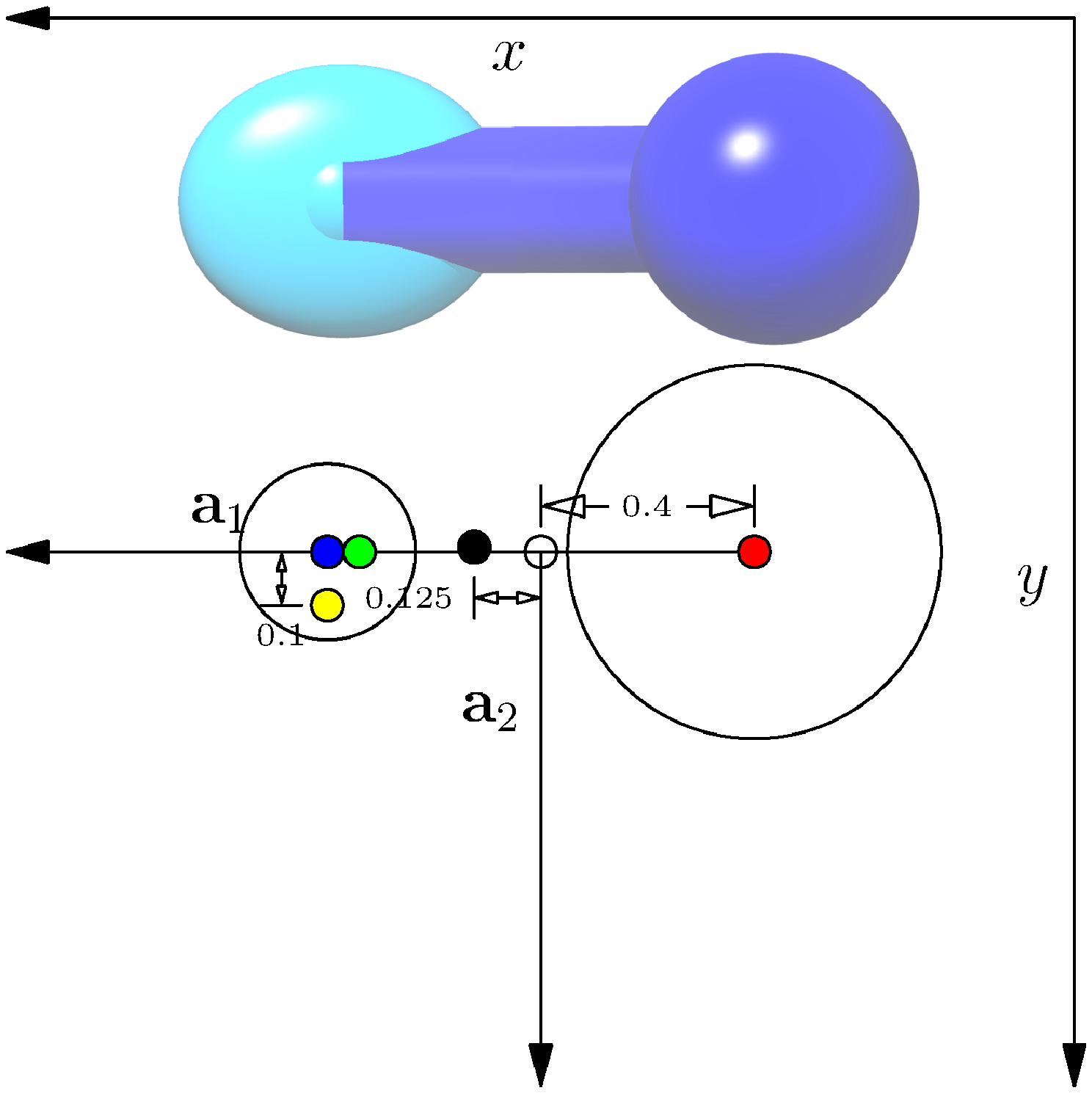}
\centering
\caption{A schematic representation of the nucleotides as represented by the oxRNA model. The red circle indicates the backbone site, the blue circle is the hydrogen-bonding site and the green circle is the coaxial stacking site. The yellow circle is the $3^{\prime}$-stacking site, and the black circle is the $5^{\prime}$ stacking site. The unfilled circle from which the $\mathbf{a}_3$, $\mathbf{a}_2$ and $\mathbf{a}_1$ vectors originate is the center of mass. All distances are given in a unit system where 1 distance unit  $= 8.4\, \angstrom$. The left image shows the projection of a single nucleotide where the $\mathbf{a}_2$ vector is pointing towards the reader, and the image on the right shows a projection where the $\mathbf{a}_3$ vector is pointing towards the reader. For comparison, we also show the schematic representation of the nucleotide that is used in producing pictures of oxRNA configurations. The backbone site is represented by a sphere because of the isotropic nature of the its interactions, whereas the base is represented by an ellipsoid whose principal axes are parallel to $\mathbf{a}_1$,  $\mathbf{a}_2$ and $\mathbf{a}_3$ respectively.}
\label{fig_nuc}
\end{figure}

The smoothed functions used in the potentials have the following form:
\begin{itemize}
\item The radial part of the stacking and hydrogen-bonding potentials:
\begin{equation}
f_1(r,\epsilon,d,r^0,r^c,r^{low},r^{high}) = \begin{cases}
	V_{\rm Morse}(r, \epsilon, r^0, d) - V_{\rm Morse}(r^{c}, \epsilon, r^0, d)) & \text{if $ r^{low} < r < r^{high} $},\\
	\epsilon V_{\rm smooth} (r, b^{low}, r^{c,low}) & \text{if $r^{c,low} < r < r^{low}$},\\
	\epsilon V_{\rm smooth} (r, b^{high}, r^{c,high}) & \text{if $r^{high} < r < r^{c,high}$},\\
	0 & \text{otherwise}.
	\end{cases} 
\end{equation}
\item The radial part of the cross-stacking and coaxial stacking potentials:
\begin{equation}
f_2(r,k,r^0,r^c,r^{low},r^{high}) = \begin{cases}
	V_{\rm harm}(r, k, r^0) - V_{\rm harm}(r^{c}, k, r^0) & \text{if $ r^{low} < r < r^{high} $},\\
	k V_{\rm smooth} (r, b^{low}, r^{c,low}) & \text{if $r^{c,low} < r < r^{low}$},\\
	k V_{\rm smooth} (r, b^{high}, r^{c,high}) & \text{if $r^{high} < r < r^{c,high}$},\\
	0 & \text{otherwise}.
	\end{cases} 
\end{equation}
\item The radial part of the excluded volume potential:
\begin{equation}
f_3(r,\epsilon,\sigma,r^{\star}) = \begin{cases}
	V_{\rm LJ}(r, \epsilon, \sigma) & \text{if $r < r^{\star} $},\\
	\epsilon V_{\rm smooth} (r, b, r^c) & \text{if $r^{\star} < r < r^c$},\\
	0 & \text{otherwise}.
	\end{cases} 
\end{equation}
\item The angular modulation factor used in stacking, hydrogen-bonding, cross-stacking and coaxial stacking:
\begin{equation}
f_4(\theta,a,\theta^0,\Delta \theta^{\star}) = \begin{cases}
	V_{\rm mod}(\theta, a, \theta^0)  & \text{if $ \theta^0 - \Delta \theta^{ \star} < \theta < \theta^0 + \Delta \theta^{\star} $},\\
	V_{\rm smooth} (\theta, b, \theta^0 - \Delta \theta^c) & \text{if $\theta^0 - \Delta \theta^c < \theta < \theta^0-\Delta \theta^{\star}$},\\
	V_{\rm smooth} (\theta, b, \theta^0 + \Delta \theta^c) & \text{if $\theta^0 + \Delta \theta^{\star} < \theta < \theta^0+\Delta \theta^c$},\\
	0 & \text{otherwise}.
	\end{cases} 
\end{equation}
\item Another modulating term which is used to impose right-handedness:
\begin{equation}
f_5(x,a,x^{\star}) = \begin{cases}
	1 & \text{if $x > 0$},\\
	V_{\rm mod} (x, a, 0) & \text{if $x^{\star} < x < 0$},\\
	V_{\rm smooth} (x, b, x^c) & \text{if $x^c< x <x^{\star}$},\\
	0 & \text{otherwise}.
	\end{cases} 
\end{equation}
\end{itemize}

We note that for given parameters of the main part of the smoothed functions (for example, $\epsilon$, $r_0$, $d$ and $r_c$ for the $V_{\rm Morse}$ part of $f_1$), the parameters of the smoothed cutoff regions ($b^{low}$, $b^{high}$, $r^{c,low}$, $r^{c,high}$ for  $f_1$) are uniquely determined by ensuring differentiability of the function at the boundaries ($r^{low}$ and $r^{high}$ for $f_1$) and thus they are not explicitly listed in the function arguments and are not provided in the tables of values of model parameters. 

The potentials are then 

\begin{equation}
 V_{\rm backbone} = V_{\rm FENE}(\delta r_{\rm backbone}, \epsilon_{\rm backbone}, \delta r^0_{\rm backbone}, \Delta_{\rm backbone}).
\end{equation}

\begin{equation}
\begin{array}{c c c}
V_{\rm stack} \left(i,j \right) & =  & \eta(i,j)( 1 + \kappa \, k_{\rm B} T) f_1(\delta r_{\rm stack}, \epsilon_{\rm stack}, d_{\rm stack}, \delta r^0_{\rm stack}, \delta r_{\rm stack}^{c},  \delta r^{low}_{\rm stack}, \delta r^{high}_{\rm stack}) \\
 & \times & f_4(\theta_{5^\prime}, a_{\rm stack,5}, \theta^0_{\rm stack,5}, \Delta \theta^{\star}_{\rm stack,5}) \,\,
 f_4(\theta_{6^\prime}, a_{\rm stack,6}, \theta^0_{\rm stack,6}, \Delta \theta^{\star}_{\rm stack,6}) \\
 & \times & f_4(\theta_{9}, a_{\rm stack,9}, \theta^0_{\rm stack,9}, \Delta \theta^{\star}_{\rm stack,9}) \,\,
 f_4(\theta_{10}, a_{\rm stack,10}, \theta^0_{\rm stack,10}, \Delta \theta^{\star}_{\rm stack,10}) \\
&  \times & f_5(\cos(\phi_1), a_{\rm stack,1},\cos(\phi_1)^{\star}_{\rm stack})  \,\,
f_5(\cos(\phi_2), a_{\rm stack,2}, \cos(\phi_2)^{\star}_{\rm stack}).
\end{array}
\end{equation}

\begin{equation}
\begin{array}{c c c}
V_{\rm H.B.}(i,j) & =  & \alpha(i,j) f_1(\delta r_{\rm HB}, \epsilon_{\rm HB}, d_{\rm HB}, \delta r^0_{\rm HB}, \delta r^{c}_{\rm HB}, \delta r^{low}_{\rm HB}, \delta r^{high}_{\rm HB})\\ 
 & \times &  f_4(\theta_1, a_{\rm HB,1}, \theta^0_{\rm HB,1}, \Delta \theta^{\star}_{\rm HB,1}) \,\,
  f_4(\theta_2, a_{\rm HB,2}, \theta^0_{\rm HB,2}, \Delta \theta^{\star}_{\rm HB,2})\\
 & \times &    f_4(\theta_3, a_{\rm HB,3}, \theta^0_{\rm HB,3}, \Delta \theta^{\star}_{\rm HB,3}) \,\,
 f_4(\theta_4, a_{\rm HB,4}, \theta^0_{\rm HB,4}, \Delta \theta^{\star}_{\rm HB,4})  \\
 & \times & f_4(\theta_7, a_{\rm HB,7}, \theta^0_{\rm HB,7}, \Delta \theta^{\star}_{\rm HB,7})  \,\,
 f_4(\theta_8, a_{\rm HB,8}, \theta^0_{\rm HB,8}, \Delta \theta^{\star}_{\rm HB,8}). \\
\end{array}
\end{equation}

\begin{equation}
\begin{array}{c c c}
 V_{\rm cross~st.}  & = & \gamma f_2(\delta r_{\rm HB}, k_{\rm cross}, \delta r^0_{\rm cross}, \delta r^{c}_{\rm cross}, \delta r^{low}_{\rm cross}, \delta r^{high}_{\rm cross})
 f_4(\theta_1, a_{\rm cross,1}, \theta^0_{\rm cross,1}, \Delta \theta^{\star}_{\rm cross,1})\\
   & \times &  f_4(\theta_2, a_{\rm cross,2}, \theta^0_{\rm cross,2}, \Delta \theta^{\star}_{\rm cross,2}) \,\,
   f_4(\theta_3, a_{\rm cross,3}, \theta^0_{\rm cross,3}, \Delta \theta^{\star}_{\rm cross,3}) \\
 & \times & \left(f_4(\theta_7, a_{\rm cross,7}, \theta^0_{\rm cross,7}, \Delta \theta^{\star}_{\rm cross,7}) + f_4(\pi-\theta_7, a_{\rm cross,7}, \theta^0_{\rm cross,7}, \Delta \theta^{\star}_{\rm cross,7}) \right)\\
 & \times & \left( f_4(\theta_8 ,a_{\rm cross,8}, \theta^0_{\rm cross,8}, \Delta \theta^{\star}_{\rm cross,8}) + f_4(\pi-\theta_8, a_{\rm cross,8}, \theta^0_{\rm cross,8}, \Delta \theta^{\star}_{\rm cross,8})  \right).\\
\end{array}
\end{equation}

\begin{equation}
\begin{array}{c c c}
 V_{\rm coaxial\,st.}  & = & \mu f_2(\delta r_{\rm coaxial\,st.}, k_{\rm coax}, \delta r^0_{\rm coax}, \delta r^{c}_{\rm coax},  \delta r^{low}_{\rm coax}, \delta r^{high}_{\rm coax})\,\,
 f_4(\theta_4, a_{\rm coax,4}, \theta^0_{\rm coax,4}, \Delta \theta^{\star}_{\rm coax,4})\\

   & \times & \left( f_4(\theta_1, a_{\rm coax,1}, \theta^0_{\rm coax,1}, \Delta \theta^{\star}_{\rm coax,1})  +  f_4(2 \pi - \theta_1, a_{\rm coax,1}, \theta^0_{\rm coax,1}, \Delta \theta^{\star}_{\rm coax,1}) \right) \\
 & \times & \left(f_4(\theta_5, a_{\rm coax,5}, \theta^0_{\rm coax,5}, \Delta \theta^{\star}_{\rm coax,5}) + f_4(\pi-\theta_5, a_{\rm coax,5}, \theta^0_{\rm coax,5}, \Delta \theta^{\star}_{\rm coax,5}) \right)\\
 & \times & \left( f_4(\theta_6 ,a_{\rm coax,6}, \theta^0_{\rm coax,6}, \Delta \theta^{\star}_{\rm coax,6}) + f_4(\pi-\theta_6, a_{\rm coax,6}, \theta^0_{\rm coax,6}, \Delta \theta^{\star}_{\rm coax,6})  \right)\\
 &  \times & f_5(\cos(\phi_3), a_{\rm coax,3^\prime},\cos(\phi_3)^{\star}_{\rm coax}) \,\,
f_5(\cos(\phi_4), a_{\rm coax,4^\prime}, \cos(\phi_4)^{\star}_{\rm coax}). 
\end{array}
\end{equation}

\begin{equation}
\begin{array}{c c c}
 V_{\rm exc} &=& f_3(\delta r_{\rm backbone},\epsilon_{\rm exc}, \sigma_{\rm backbone},\delta r_{\rm backbone}^{\star} ) + f_3(\delta r_{\rm HB},\epsilon_{\rm exc}, \sigma_{\rm base},\delta r_{\rm base}^{\star} ) \\    
             &+& f_3(\delta r_{\rm back-base},\epsilon_{\rm back-base}, \sigma_{\rm back-base},\delta r_{\rm back-base}^{\star} ) \\
             &+& f_3(\delta r_{\rm base-back},\epsilon_{\rm back-base}, \sigma_{\rm back-base},\delta r_{\rm back-base}^{\star} ) \\
 \end{array}
\end{equation}

The excluded volume interaction between bonded neighbors, $V_{\rm exc}^{\prime}$, is the same as $V_{\rm exc}$ with the exception that it does not include the first term which depends on 
$\delta r_{\rm backbone}$, because the neighbors already interact with the FENE potential through the $V_{\rm backbone}$ interaction that ensures that the backbone sites do not come too close.

The parameters of the interaction potentials are specified in tables \ref{parameters} and \ref{parametersb}. 

\begin{table*}
\begin{center}
\begin{tabular}{c  c  c  c }
\hline
& & &\\
{\bf Interaction} & \multicolumn{3}{c}{  \bf Parameters }\\\
\\ \hline \hline
& & &\\
 \multicolumn{4}{c} {backbone spring: $V_{\rm backbone}$}\\
$V_{\rm FENE}(\delta r_{\rm backbone})$ & $\epsilon_{\rm backbone} = 2$ & $\Delta_{\rm backbone}= 0.25$ & $\delta r^0_{\rm backbone} = 0.76$ \\ 
\hline
& & & \\
 \multicolumn{4}{c} {hydrogen bonding: $V_{\rm H.B.}$ } \\
$f_1(\delta r_{\rm HB})$& $\epsilon_{\rm HB} = 1.0$ & $d_{\rm HB} = 8$ & $\delta r^0_{\rm HB} = 0.4$ \\
 $\alpha^{\rm avg} = 0.87$ & $ \alpha({\rm A,U}) = 0.82$ & $ \alpha({\rm G,U}) = 0.51$ & $ \alpha({\rm G,C}) = 1.06$ \\
  & $\delta r^c_{\rm HB}=0.75$ & $\delta r^{low}_{\rm HB} = 0.34$  & $\delta r^{high}_{\rm HB} = 0.70$ \\
$f_4(\theta_1)$ & $a_{\rm HB,1} = 1.50$ & $\theta^0_{\rm HB,1} = 0$ & $\Delta \theta^{\star}_{\rm HB,1} = 0.70$\\
$f_4(\theta_2)$ & $a_{\rm HB,2}= 1.50$ & $\theta^0_{\rm HB,2} = 0$ & $\Delta \theta^{\star}_{\rm HB,2}= 0.70$\\
$f_4(\theta_3)$ & $a_{\rm HB,3}= 1.50$ & $\theta^0_{\rm HB,3} = 0$ & $\Delta \theta^{\star}_{\rm HB,3} = 0.70$\\
$f_4(\theta_4)$& $a_{\rm HB,4} = 0.46$ & $\theta^0_{\rm HB,4} =  \pi$ & $\Delta \theta^{\star}_{\rm HB,4} = 0.70$\\
$f_4(\theta_7)$ & $a_{\rm HB,7}= 4.00$ & $\theta^0_{\rm HB,7} = \pi/2$ & $\Delta \theta^{\star}_{\rm HB,7} = 0.45$\\
$f_4(\theta_8)$ & $a_{\rm HB,8} = 4.00$ & $\theta^0_{\rm HB,8} =  \pi/2$ & $\Delta \theta^{\star}_{\rm HB,8}= 0.45$\\
\hline
& & & \\
 \multicolumn{4}{c} {stacking: $V_{\rm stack}$} \\
$f_1(\delta r_{\rm stack})$& $\epsilon_{\rm stack} =  1.0$ & $d_{\rm stack} = 6$ & $\delta r^0_{\rm stack} = 0.43$ \\
& $\delta r^c_{\rm stack}=0.93$ & $\delta r^{low}_{\rm stack} = 0.35$  &  $\delta r^{high}_{\rm stack} =  0.78$ \\
 & $\eta^{\rm avg} = 1.402$ & $\kappa = 1.9756 $ &\\
$\eta({\rm G,C}) = 1.276 $ & $\eta({\rm C,G}) = 1.603 $ & $\eta({\rm G,G}) = 1.494 $ & $\eta({\rm C,C}) = 1.473 $ \\
$\eta({\rm G,A}) = 1.621 $ & $\eta({\rm U,C}) = 1.167 $ & $\eta({\rm A,G}) = 1.394 $ & $\eta({\rm C,U}) = 1.471 $ \\ 
$\eta({\rm U,G}) = 1.286 $ & $\eta({\rm C,A}) = 1.583 $ & $\eta({\rm G,U}) = 1.571 $ & $\eta({\rm A,C}) = 1.210 $  \\
$\eta({\rm A,U}) = 1.385 $ & $\eta({\rm U,A}) = 1.246 $ & $\eta({\rm A,A}) = 1.316 $ & $\eta({\rm U,U}) = 1.175 $ \\

$f_4(\theta_{5^\prime})$ & $a_{\rm stack,5}  = 0.90$ & $\theta^0_{\rm stack,5}  = 0$ & $\Delta \theta^{\star}_{\rm stack,5}  = 0.95$ \\
 $f_4(\theta_{6^\prime})$ & $a_{\rm stack,6}  = 0.90$ & $\theta^0_{\rm stack,6}  = 0$ & $\Delta \theta^{\star}_{\rm stack,6} = 0.95$  \\
 $f_4(\theta_{9})$ & $a_{\rm stack,9}  = 1.3$ & $\theta^0_{\rm stack,9}  = 0$ & $\Delta \theta^{\star}_{\rm stack,9}  =  0.8$ \\
 $f_4(\theta_{10})$ & $a_{\rm stack,10}  = 1.3$ & $\theta^0_{\rm stack,10}  = 0$ & $\Delta \theta^{\star}_{\rm stack,10} = 0.8$  \\
$f_5(\cos(\phi_1))$ & $a_{\rm stack,1} =2.00$ &   $\cos(\phi_1)^{\star}_{\rm stack}= 0.65$  \\ 
$f_5(\cos(\phi_2))$ & $a_{\rm stack,2} =2.00$ &  $\cos(\phi_2)^{\star}_{\rm stack}= 0.65$  \\
\hline
& & & \\
\multicolumn{4}{c} {excluded volume: $V_{\rm exc}$} \\
$f_3(\delta r_{\rm backbone})$ & $\epsilon_{\rm exc} = 2.00$ & $\sigma_{\rm backbone} = 0.70$ & $\delta r_{\rm backbone}^{\star} =0.675$ \\
$f_3(\delta r_{\rm HB})$ & $\epsilon_{\rm exc} = 2.00$ & $\sigma_{\rm base} = 0.33$ & $\delta r_{\rm base}^{\star} =0.32$ \\
$f_3(\delta r_{\rm back-base})$ & $\epsilon_{\rm exc} = 2.00$ & $\sigma_{\rm back-base} = 0.515$ & $\delta r_{\rm back-base}^{\star} =0.50$ \\
$f_3(\delta r_{\rm base-back})$ & $\epsilon_{\rm exc} = 2.00$ & $\sigma_{\rm back-base} = 0.515$ & $\delta r_{\rm back-base}^{\star} =0.50$ \\
\hline
\end{tabular}
\caption{Parameter values in the model. In this table, all energies and lengths are in terms of the simulation units of energy and distance. When more than one function is listed for an interaction, the total interaction is a product of all the terms with the exception of $V_{\rm exc}$, which is a sum of the respective terms.}
\label{parameters}
\end{center}
\end{table*}

\begin{table*}
\begin{center}
\begin{tabular}{c  c  c  c }
\hline
& & &\\
{\bf Interaction} & \multicolumn{3}{c}{  \bf Parameters }\\\
\\ \hline \hline
& & &\\
 \multicolumn{4}{c} {cross-stacking: $V_{\rm cross.~st.}$}\\
$f_2(\delta r_{\rm HB})$ & $k_{\rm cross} = 1.0$ & $r^0_{\rm cross} =  0.5$ & $\delta r^c_{\rm cross}=0.6$ \\
$\gamma =  59.96 $& $\delta r^{low}_{\rm cross} = 0.42$ & $\delta r^{high}_{\rm cross} =  0.58$ & \\
$f_4(\theta_1)$ & $a_{\rm cross,1} = 2.25$ & $\theta^0_{\rm cross,1} =  0.505$ & $\Delta \theta^{\star}_{\rm cross,1} = 0.58$\\
$f_4(\theta_2)$ & $a_{\rm cross,2}= 1.70$ & $\theta^0_{\rm cross,2} =  1.266$ & $\Delta \theta^{\star}_{\rm cross,2}= 0.68$\\
$f_4(\theta_3)$ & $a_{\rm cross,3}= 1.70$ & $\theta^0_{\rm cross,3} =  1.266$ & $\Delta \theta^{\star}_{\rm cross,3} = 0.68$\\
$f_4(\theta_7) + f_4(\pi - \theta_7)$ & $a_{\rm cross,7}= 1.70$ & $\theta^0_{\rm cross,7} = 0.309$ & $\Delta \theta^{\star}_{\rm cross,7} = 0.68$\\
$f_4(\theta_8) + f_4(\pi - \theta_8) $ & $a_{\rm cross,8} = 1.70$ & $\theta^0_{\rm cross,8} =  0.309$ & $\Delta \theta^{\star}_{\rm cross,8}= 0.68$\\
\hline
& & & \\
 \multicolumn{4}{c} {coaxial stacking: $V_{\rm coaxial\, st.}$} \\
$ f_2(\delta r_{\rm coax})$ & $k_{\rm coax} = 1.0$  & $\delta r_{\rm coax}^0 = 0.5$  & $\delta r_{\rm coax}^c = 0.6$  \\
$\mu =  80.0$ & $\delta r_{\rm coax}^{low} =0.42$ & $\delta r_{\rm coax}^{high} = 0.58$\\
$f_4(\theta_1) + f_4(2\pi-\theta_1)$ & $a_{\rm coax,1} = 2.00$ & $\theta^0_{\rm coax,1} = 2.592$ & $\Delta \theta^{\star}_{\rm coax,1} = 0.65$   \\
$f_4(\theta_4)$ & $a_{\rm coax,4} = 1.30$ & $\theta_{\rm coax,4}^0 = 0.151$ & $\Delta \theta^{\star}_{\rm coax,4}= 0.8$\\
$f_4(\theta_5)+f_4(\pi-\theta_5)$ & $a_{\rm coax,5} = 0.90$ & $\theta_{\rm coax,5}^0 = 0.685$ & $\Delta \theta^{\star}_{\rm coax,5} = 0.95$ \\
$f_4(\theta_6)+f_4(\pi-\theta_6)$ & $a_{\rm coax,6} = 0.90$ & $\theta_{\rm coax,6}^0 = 2.523$ & $\Delta \theta^{\star}_{\rm coax,6}= 0.95$ \\
$f_5(\cos(\phi_3))$ & $a_{\rm coax,3^\prime}=2.00$ &   $\cos(\phi_3)_{\rm coax}^{\star}= -0.65$  \\ 
$f_5(\cos(\phi_4))$ & $a_{\rm coax,4^\prime}=2.00$ &  $\cos(\phi_4)_{\rm coax}^{\star}= -0.65$  \\ 
\hline
\end{tabular}
\caption{Further model parameters.}
\label{parametersb}
\end{center}
\end{table*}

\end{widetext}

\bibliographystyle{aip}
\bibliography{rna_biblio2}
\end{document}